\documentclass[journal]{IEEEtran}
\usepackage{cite}
\usepackage{amsmath,amssymb,amsfonts}
\usepackage{algorithmic}
\usepackage{graphicx}
\usepackage{amssymb,comment}
\usepackage{algorithmic}
\usepackage{algorithm}
\usepackage{caption} 
\usepackage{subfig}
\usepackage{subcaption}
\usepackage{scrextend}
\usepackage{verbatim}
\usepackage{color}
\usepackage{enumitem,epstopdf}

\newtheorem{lemma}{\bf Lemma}
\newtheorem{remark}{\bf Remark}

\newcommand{\bs}[1]{\boldsymbol{#1}}
\newcommand{\mbf}[1]{\mathbf{#1}}
\newcommand{\ib}[1]{\in\mathbb{#1}}
\newcommand{\ic}[1]{\in\mathcal{#1}}
\newcommand{\ca}[1]{\mathcal{#1}}

\usepackage{setspace}

\begin{document}
	\title{Movable Antennas Meet Intelligent Reflecting Surface: Friends or Foes?}
	\author{Xin~Wei,~\IEEEmembership{Graduate Student Member,~IEEE,}
		Weidong~Mei,~\IEEEmembership{Member,~IEEE,}
		Qingqing~Wu,~\IEEEmembership{Senior Member,~IEEE,}
		Qiaoran~Jia,
		Boyu Ning,~\IEEEmembership{Member,~IEEE,}
		Zhi~Chen,~\IEEEmembership{Senior Member,~IEEE,}
		and~Jun~Fang,~\IEEEmembership{Senior Member,~IEEE}\vspace{-6pt}
		\thanks{This paper has been presented in part at the IEEE Wireless Communications and Networking Conference (WCNC), Milan, Italy, 2025 \cite{wei2025movable}.
			
		Xin Wei, Weidong Mei, Boyu Ning, Zhi Chen, and Jun Fang are with the National Key Laboratory of Wireless Communications, University of Electronic Science and Technology of China, Chengdu 611731, China (e-mail: xinwei@std.uestc.edu.cn, wmei@uestc.edu.cn, boydning@outlook.com, chenzhi@uestc.edu.cn, junfang@uestc.edu.cn).
		
		Qingqing Wu is with the Department of Electronic Engineering, Shanghai Jiao Tong University, Shanghai 200240, China (email: qingqingwu@sjtu.edu.cn).
		
		Qiaoran Jia is with the Glasgow College, University of Electronic Science and Technology of China, Chengdu 611731, China (email:  2022190901018@std.uestc.edu.cn).
		}
	}
	
	\maketitle
	\begin{abstract}
		Movable antenna (MA) and intelligent reflecting surface (IRS) are considered promising technologies for the next-generation wireless communication systems due to their shared capabilities of reconfiguring and improving wireless channel conditions. This, however, raises a fundamental question: Does the performance gain of MAs over conventional fixed-position antennas (FPAs) still exist in the presence of the IRS passive beamforming? To answer this question, we investigate in this paper an IRS-assisted multi-user multiple-input single-output (MISO) MA system, where a multi-MA base station (BS) transmits to multiple single-FPA users. We formulate a sum-rate maximization problem by jointly optimizing the active/passive beamforming of the BS/IRS and the MA positions within a one-dimensional transmit region, which is challenging to be optimally solved. To drive essential insights, we first study a simplified case with a single user. Then, we analyze the performance gain of MAs over FPAs in the light-of-sight (LoS) BS-IRS channel and derive the conditions under which this gain becomes more or less significant. In addition, we propose an alternating optimization (AO) algorithm to solve the signal-to-noise ratio (SNR) maximization problem in the single-user case by combining the block coordinate descent (BCD) method and the graph-based method. For the general multi-user case, our performance analysis unveils that the performance gain of MAs over FPAs diminishes with typical transmit precoding strategies at the BS under certain conditions. We also propose a high-quality suboptimal solution to the sum-rate maximization problem by applying the AO algorithm that combines the weighted minimum mean square error (WMMSE) algorithm, manifold optimization method and discrete sampling method. Numerical results validate our theoretical analyses and demonstrate that the performance gain of MAs over FPAs may be reduced if the IRS passive beamforming is optimized.\vspace{-3pt}
	\end{abstract}
	\begin{IEEEkeywords}
		Movable antennas, intelligent reflecting surfaces, near-field communications, performance analysis, alternating optimization, graph theory.\vspace{-6pt}
	\end{IEEEkeywords}
	
	\section{Introduction}
	Intelligent reflective surface (IRS) has been deemed as a promising technology for future sixth-generation (6G) wireless networks, owing to its channel reconfiguration capability, low power consumption, and low-cost deployment \cite{wu2024intelligent}. Specifically, an IRS consists of a large array of passive reflecting elements that can independently adjust the amplitude and/or phase of incident signals. By properly deploying the IRSs and adjusting their reflections, IRSs can jointly alter the strength/direction of their reflected signals for achieving various purposes, e.g., interference suppression \cite{wu2020towards,pan2020multicell}, multi-reflection coverage extension \cite{mei2022intelligent}, spatial multiplexing \cite{zhang2020capacity,ning2021terahertz}, target sensing \cite{shao2024intelligent}, among others.
	
	On the other hand, movable antenna (MA) (also known as fluid antenna system (FAS) for antenna position adjustment \cite{wong2021fluid}) has drawn great interest in academia and industry recently. Similar to the IRS, MA can reconfigure wireless channel conditions by enabling multiple antennas to be flexibly moved within a confined region at the transmitter or receiver \cite{ning2024movable,zhu2024movable,wang2024empowered}. The antenna position optimization for MAs has been investigated in various scenarios, such as flexible beamforming \cite{ma2024multi,zhu2023movable,wang2024flexible}, physical-layer security \cite{tang2024secure,hu2024secure}, multiple-input multiple-output (MIMO) system \cite{ma2024mimo}, over-the-air computation \cite{li2025over}, cognitive radio \cite{wei2024joint}, integrated sensing and communication (ISAC) \cite{wang2024fluid,lyu2025movable}, etc. Particularly, most of the existing works applied the gradient-based algorithms, e.g., the successive convex approximation (SCA) technique \cite{ma2024multi,tang2024secure,ma2024mimo,shao2025movable} and gradient ascend/descent method \cite{hu2024secure,li2024movable}, to optimize the antenna positions. In addition, some other works proposed a variety of evolutionary algorithms to optimize the antenna positions, such as the particle swarm optimization (PSO) algorithm and its derivative algorithms \cite{li2025over,xiao2024multiuser,shao2024exploiting}. Furthermore, the deep learning-based method was introduced in \cite{kang2024deep} for MA-assisted multicasting, where the feedforward neural networks (FNNs) were utilized to extract useful features from the input direction information and jointly optimize the beamforming and antenna positions. Instead of optimizing the antenna positions in a continuous space, the authors of \cite{mei2024movable,wei2024joint,mei2024posistion} discretized the movable region into a multitude of sampling points and proposed a graph-based method to select an optimal subset of sampling points for maximizing the communication rates. However, all of the above position optimization algorithms rely on global channel state information (CSI) at any position within the transmit/receive movable region. To facilitate the practical implementation, the authors of \cite{zeng2025position} proposed a CSI-free MA position optimization approach via the zero-order gradient approximation method.
	
 	To reap the potential combined gain of IRS and MA, some works have investigated the integration of MAs into the IRS-assisted wireless system \cite{rostami2024performance,lai2024fas,chen2024low,sun2024optimization,xiao2024throughput,zhang2024enhancement}. For example, the authors of \cite{rostami2024performance} analyzed the outage probability and delay outage rate of an IRS-aided single-input single-output (SISO) FAS. To simplify the expressions of outage probability and reduce the computational complexity, the authors of \cite{lai2024fas} adopted a block-correlation channel model and analyzed the impact of the number of fluid antenna ports on the system performance. A low-complexity beamforming design for IRS-assisted multiple-input single-output (MISO) FAS was proposed in \cite{chen2024low}, which only requires statistical CSI. The sum-rate maximization problem for IRS-assisted MA systems was investigated in \cite{sun2024optimization,xiao2024throughput,zhang2024enhancement}. In particular, the authors of \cite{sun2024optimization} studied a multi-MA BS serving multiple users each with a single fixed-position antenna (FPA), in the presence of an IRS with different settings of its reflection coefficients. The authors of \cite{xiao2024throughput} studied an IRS-assisted wireless energy transfer system in an Internet of Things (IoT) network with multiple MA-enabled IoT devices. Different from the passive IRS considered in \cite{rostami2024performance,chen2024low,sun2024optimization,xiao2024throughput,lai2024fas}, an active IRS was introduced in \cite{zhang2024enhancement} for further performance enhancement, where a fractional programming-based genetic algorithm was proposed to solve the non-convex sum-rate maximization problem. Nonetheless, most of the above works only focus on the joint impacts of MA and IRS on the system performance via algorithm designs, without delving into the potential mutual interactions between them. In addition, most of them only consider a single-MA BS \cite{rostami2024performance,lai2024fas} or single-user \cite{rostami2024performance,lai2024fas,chen2024low} system setup. Due to the comparable capability of MA and IRS for channel reconfiguration, it remains an open problem whether the performance gain of MAs over FPAs still exists in the presence of the IRS's passive beamforming. 	
 	
 	To resolve this problem, in this paper, we conduct performance analysis and optimization for an IRS-assisted multi-user MISO MA system, where a BS equipped with multiple MAs transmits to multiple single-FPA users with the aid of an IRS, as shown in Fig. 1. The main contributions of this paper are summarized as follows.
 	\begin{itemize}
 		\item We formulate a sum-rate maximization problem for the considered system, aiming to jointly optimize the BS's transmit precoding, the IRS's passive beamforming and the positions of MAs within a one-dimensional (1D) transmit region. Since this optimization problem is non-convex and difficult to be optimally solved, we first consider a simplified case with a single user and aim to maximize its received signal-to-noise ratio (SNR). To drive useful insights, we conduct theoretical analysis of the performance gain of MAs over conventional FPAs under a general non-uniform spherical wave (NUSW)-based line-of-sight (LoS) channel model between the BS and the IRS. It is unveiled that with the optimal IRS passive beamforming, the performance gain of MAs over FPAs may vanish in the case of a single-MA BS or far-field BS-IRS channel. Moreover, we derive the conditions under which the performance gain becomes more or less significant. For the SNR maximization problem, we decompose it into two subproblems and solve them alternately by adopting the block coordinate descent (BCD) and graph-based algorithms, respectively.
 		
 		\item For the general multi-user case, we first analyze the performance gain of MAs with various transmit precoding schemes at the BS, such as zero-forcing (ZF) and minimum mean square error (MMSE). It is revealed that in the far-field LoS BS-IRS channel, the MAs cannot yield any performance gain over FPAs with these typical precoding schemes, which is in sharp contrast with the case without IRS. Next, to solve the sum-rate maximization problem, we propose an alternating optimization (AO) algorithm by solving three sub-problems alternately using the weighted MMSE (WMMSE) algorithm, the conjugate gradient (CG)-based manifold method, and the discrete-sampling-based method, respectively. Finally, simulation results are presented to validate our analysis and demonstrate the efficacy of our proposed AO algorithm in both single- and multi-user setups. It is also shown that the performance gain of MAs may be reduced by the optimized IRS passive beamforming.
 	\end{itemize}
	
	\begin{figure}[!t]
		\centering
		\captionsetup{justification=raggedright,singlelinecheck=false}
		\centerline{\includegraphics[width=0.5\textwidth]{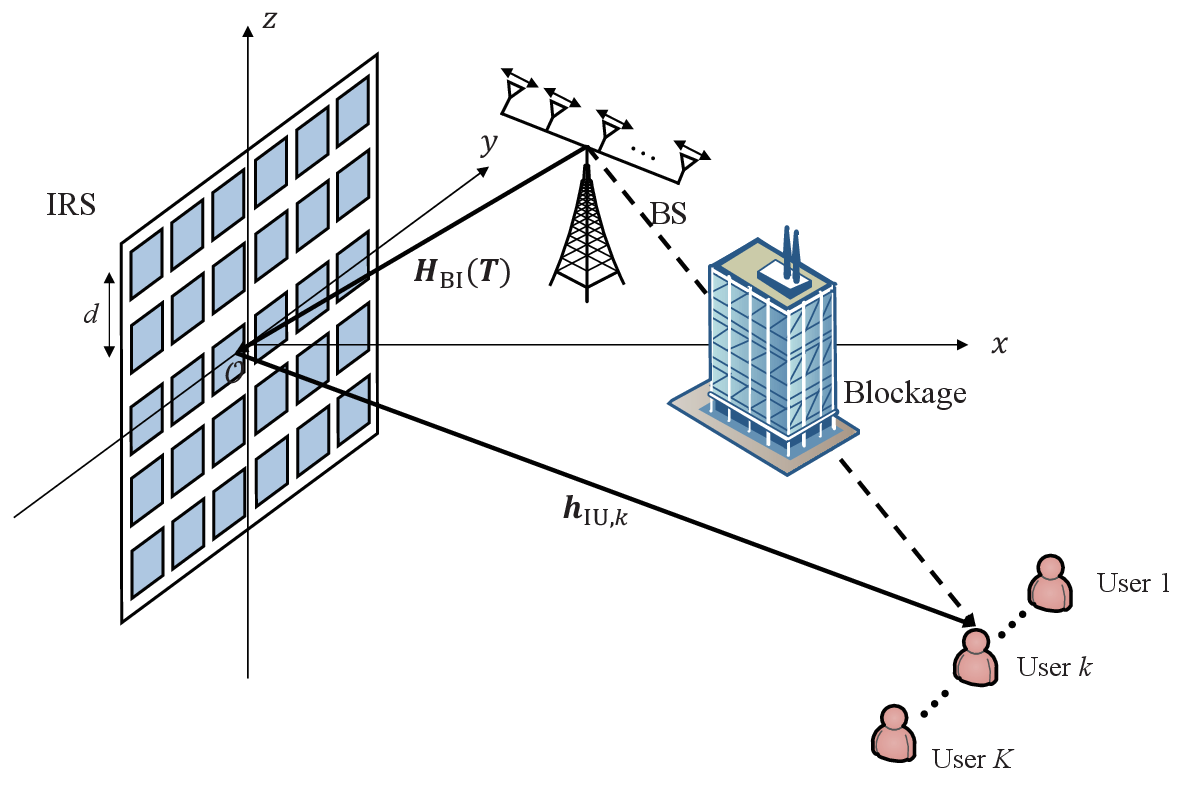}}
		\captionsetup{font=footnotesize}
		\vspace{-9pt}
		\caption{IRS-aided multi-user MISO MA system.}
		\label{Fig_SysModel}
		\vspace{-15pt}
	\end{figure}
	
	It is worth noting that in addition to integrating MAs into an IRS-assisted communication system, another branch of works (see e.g., \cite{hu2024intelligent,zhang2024wireless,liu2024uav}) propose a new architecture of IRSs with movable/rotatable reflecting elements, which is different from the system setup considered in this paper.
	
	The rest of this paper is organized as follows. Section II presents the system model. Section III presents the performance analysis and our proposed optimization algorithms for the single-user case, while Section IV presents those for the general multi-user case. Section V presents the numerical results. Finally, Section VI concludes this paper and discusses future directions.
	
	{\it Notations:} $a$, $\bs{a}$, $\bs{A}$, and $\ca{A}$ denote a scalar, a vector, a matrix and a set respectively. For a complex number $a$, $\angle a$, $|a|$, and $a^*$ denote its phase, amplitude and conjugate, respectively. $(\cdot)^T$, $(\cdot)^H$, and $(\cdot)^{-1}$ denote the transpose, conjugate transpose and inverse of a matrix, respectively. $\mathbb{R}$ and $\mathbb{C}$ denote the sets of real numbers and complex numbers, respectively. $|a|$ and $||\bs{a}||_2$ denote the amplitude of a scalar $a$ and the norm of a vector $\bs{a}$, respectively. $[\bs{a}]_n$ and $[\bs{A}]_{n,m}$ denote the $n$-th entry of a vector $\bs{a}$ and the element at the $n$-th row and the $m$-th column of a matrix $\bs{A}$, respectively. The operator $\odot$ denotes the Hadamard product. The operator $\mathrm{unt}(\bs{a})$ divides each entry of $\bs{a}$ by its amplitude, i.e., $\mathrm{unt}(\bs{a})=\left[\frac{a_1}{|a_1|},\frac{a_2}{|a_2|},\cdots,\frac{a_n}{|a_n|}\right]$. $x\sim\ca{CN}(\mu,\sigma^2)$ represents that $x$ follows the circularly symmetric complex Gaussian (CSCG) distribution with the mean $\mu$ and variance $\sigma^2$, while $x\sim\ca{U}[a,b]$ represents that $x$ follows the uniform distribution between $a$ and $b$.\vspace{-6pt}
	
	\begingroup
	\allowdisplaybreaks
	\section{System Model and Problem Formulation}
	As shown in Fig. 1, we consider the downlink of a multi-user MISO wireless communication system, where an IRS is deployed to assist in the data transmission from a BS to $K$ users. Let $\mathcal{K}\triangleq\left\{1,2,\cdots,K\right\}$ denote the set of the $K$ users. To focus on the mutual interactions between the MAs and the IRS, the direct channels from the BS to all users are assumed to be negligible in this paper to eliminate their effects. We assume that the BS is equipped with $N$ ($N\ge1$) MAs, while each user is equipped with a single FPA. For convenience, we establish a global three-dimensional (3D) coordinate system, where the IRS is assumed to be located in the $yOz$-plane and centered at the origin, as shown in Fig. 1. The total number of IRS reflecting elements is denoted as $M=M_yM_z$, with $M_y$ and $M_z$ denoting the number of reflecting elements along the $y$- and $z$-axes, respectively. Without loss of generality, we assume that $M_y$ and $M_z$ are both even numbers. As such, the coordinates of the $(m_y,m_z)$-th IRS reflecting element is given by $\bs{e}_{m_y,m_z}=[0,m_yd,m_zd]^T$, where $m_y\ic{M}_y\triangleq\{0,\pm1,\pm2,\cdots,\pm\frac{M_y}{2}\}$, $m_z\ic{M}_z\triangleq\{0,\pm1,\pm2,\cdots,\pm\frac{M_z}{2}\}$. Let $\bs{\Phi}=\text{diag}\left(\bs{\varphi}\right)=\text{diag}\left([e^{j\varphi_1},e^{j\varphi_2},\cdots,e^{j\varphi_M}]^T\right)\ib{C}^{M\times M}$ denote the reflection matrix of the IRS, where $\varphi_m$ denotes the phase shift of the $m$-th reflecting element, $m\ic{M}\triangleq\{1,2,\cdots,M\}$, and $\bs{\varphi}=[e^{j\varphi_1},e^{j\varphi_2},\cdots,e^{j\varphi_M}]^T$ denotes the passive beamforming vector of the IRS.
	
	Furthermore, we assume that the MAs at the BS can be flexibly moved within a linear array\footnote{The proposed algorithms and theoretical results in this paper are also applicable to the case with a two-dimensional (2D) transmit region at the BS.} of $A$ meters (m), denoted as ${\cal C}_t$. The coordinate of its center is denoted as $\bs{q}_B=[x_B,y_B,z_B]^T$. Let $\bs{t}_n=[x_n,y_n,z_n]^T$ denote the coordinate of the $n$-th MA, $n\ic{N}\triangleq\{1,2,\cdots,N\}$, and $\bs{T}=[\bs{t}_1,\bs{t}_2,\cdots,\bs{t}_N]^T\ib{R}^{3\times N}$ denote the antenna position vector (APV) of the $N$ MAs.
	
	Let $\bs{H}_{\text{BI}}(\bs{T})\ib{C}^{M\times N}$ and $\bs{h}_{\text{IU},k}\ib{C}^{M\times1}$ denote the BS-IRS and IRS-user $k$ channels, respectively. Note that the BS-IRS channel depends on the APV, i.e., $\bs{T}$, while $\bs{h}_{\text{IU},k}$ is regardless of it. As such, the cascaded BS-IRS-user $k$ channel can be expressed as $\bs{h}_k^H(\bs{T},\bs{\Phi})=\bs{h}_{\text{IU},k}^H\bs{\Phi}\bs{H}_{\text{BI}}(\bs{T})\ib{C}^{1\times N}$, $\forall k\ic{K}$. Let $\bs{w}_k\ib{C}^{N\times 1}$ denote the BS's transmit precoding for user $k$ with $\sum_{k\ic{K}}{||\bs{w}_k||_2^2}\le P$, where $P$ is the BS's transmit power. Then, the received signal at user $k$ is given by 
	\begin{equation}\label{eqn_Signal}
		\begin{aligned}
			y_k=\bs{h}_k^H(\bs{T},\bs{\Phi})\bs{w}_kx_k+\sum_{i\ne k}{\bs{h}_k^H(\bs{T},\bs{\Phi})\bs{w}_ix_i}+n_k,
		\end{aligned}
	\end{equation}
	where $x_k$ is the transmitted data symbol for user $k$ with $\mathbb{E}[|x_k|^2]=1$, and $n_k\sim\mathcal{CN}(0,\sigma^2)$ represents the received noise at the user with $\sigma^2$ denoting its average power. Based on \eqref{eqn_Signal}, the achievable rate at user $k$ is given by
	\begin{equation}\label{eqn_SINR}
		R_k(\bs{W},\bs{\Phi},\bs{T})=\log_2\left(1+\frac{\left|\bs{h}_k^H(\bs{\Phi},\bs{T})\bs{w}_k\right|^2}{\sum_{i\ne k}{\left|\bs{h}_k^H(\bs{\Phi},\bs{T})\bs{w}_i\right|^2}+\sigma^2}\right),
	\end{equation}
	where $\bs{W}=[\bs{w}_1,\bs{w}_2,\cdots,\bs{w}_K]\ib{C}^{N\times K}$ denotes the transmit precoding matrix. Our objective is to maximize the sum-rate of the $K$ users, by jointly optimizing the BS's transmit precoding matrix $\bs{W}$, the IRS's reflection matrix $\bs{\Phi}$ and the APV $\bs{T}$. Hence, the optimization problem is formulated as\footnote{Note that the sum-rate maximization problem can be viewed as a dual of the bit-error rate (BER) minimization problem by introducing a BER-related penalty term to the SINRs within the logarithm \cite{goldsmith2005wireless}. Moreover, in the single-user case, the rate maximization becomes equivalent to the BER minimization, since the BER decreases with increasing the user's achievable rate or SNR.}
	\begin{subequations}\label{eqn_OptPrblm_P1}
		\begin{align}
			{\text{(P1)}}\quad &\underset{\bs{\Phi},\bs{T},\bs{W}}{\max}\quad R_{\mathrm{sum}}=\sum_{k\ic{K}}R_k(\bs{W},\bs{\Phi},\bs{T}) \nonumber
			\\
			\mathrm{s.t.}\quad & \varphi_m\in[0,2\pi], \forall m \ic{M}, \label{eqn_IRS_Phase_Cons} \\
			& \bs{t}_n \ic{C}_t, n \ic{N}, \label{eqn_MA_Region_Cons} \\
			& ||\bs{t}_i - \bs{t}_j||_2 \ge D_{\min}, \forall i,j\ic{N}, i\ne j, \label{eqn_MA_Coordinate_Cons}\\
			& \sum_{k\ic{K}}{||\bs{w}_k||_2^2}\le P,\label{eqn_Pw_Cons}
		\end{align}
	\end{subequations}
	where $D_{\min}$ denotes the minimum spacing between any two MAs to avoid mutual coupling. To characterize the performance limit and ease the performance analysis, we assume that all required CSI is available by applying the existing channel estimation techniques dedicated to MAs/IRSs, e.g., compressed sensing \cite{ma2023compressed,zheng2022survey}. For example, a viable solution is by equipping the IRS with a few active sensors, such that the BS-IRS channel (for all MA positions) and IRS-user channel can be separately estimated. In addition, we assume that the antenna movement delay is negligible compared to the channel coherence time, which usually holds in various slow-varying scenarios, e.g., smart homes and factories. Notably, the movement delay can be further reduced by employing advanced electronically driven solutions for MAs \cite{ning2024movable}, which also helps prevent potential inter-antenna collisions associated with physical movement.
	
	However, (P1) is a non-convex optimization problem that is challenging to solve due to the inter-MA spacing constraints (i.e., \eqref{eqn_MA_Coordinate_Cons}) and the unit-modulus constraints for IRS passive beamforming (i.e., \eqref{eqn_IRS_Phase_Cons}). To reveal essential insights, we first consider a simplified case with a single user and conduct performance analyses in the next section.
	
	\section{Single-User Case}
	In this section, we consider a special single-user scenario for (P1) to obtain some useful insights. In this scenario, maximizing the user's achievable rate is equivalent to maximizing the received SNR at the user, which is given by
	\begin{equation}\label{eqn_SNR}
		\gamma(\bs{w},\bs{\Phi},\bs{T})=\frac{P}{\sigma^2}\left|\bs{h}_{\text{IU}}^H\bs{\Phi}\bs{H}_{\text{BI}}(\bs{T})\bs{w}\right|^2,
	\end{equation}
	where $\bs{w}\ib{C}^{N\times1}$ denotes the transmit beamforming with $||\bs{w}||_2=1$, and the subscripts “$k$” of $\bs{h}_{\text{IU}}$ and $\bs{w}$ are omitted here without ambiguity. As such, (P1) reduces to the following SNR maximization problem, i.e.,
	\begin{subequations}\label{eqn_OptPrblm_P2}
		\begin{align}
			{\text{(P2)}}\quad &\underset{\bs{\Phi},\bs{T},\bs{w}}{\max}\quad\gamma(\bs{w},\bs{\Phi},\bs{T}) \nonumber
			\\
			\mathrm{s.t.}\quad & ||\bs{w}||_2=1,\,\, \eqref{eqn_IRS_Phase_Cons},\eqref{eqn_MA_Region_Cons},\eqref{eqn_MA_Coordinate_Cons}.
		\end{align}
	\end{subequations}\vspace{-25pt}
	
	\subsection{Performance Analysis}
	To facilitate our performance analyses, we first consider that the BS is equipped with a single MA, i.e., $N=1$, and characterize the maximum BS-user end-to-end channel power gain over different positions within the transmit region $\ca{C}_t$ under the optimal IRS passive beamforming. In this case, the APV reduces to a column vector, $\bs{t}\ib{R}^{3\times1}=[x_t, y_t, z_t]^T$. Moreover, we consider that the IRS can achieve a LoS-dominant channel with the BS, which usually holds in practice by carefully deploying the IRS, e.g., in the vicinity of the BS \cite{huang2022empowering}. Note that the effective length of the MA array varies with the positions of the MAs, which also alters the boundaries between the near- and far-field regions, as well as the corresponding channel models. To reduce the inaccuracy due to the channel model mismatch, we consider the maximum length of the MA array (i.e., $A$) to determine the BS-IRS channel model. Hence, the maximum Rayleigh distance between the BS and the IRS is given by \cite{zhou2015spherical,lu2023near}
	\begin{equation}\label{eqn_RayD}
		R_{\text{Ray}}=\frac{2(D_{\text{IRS}}+A)^2}{\lambda},
	\end{equation}	
	where $D_{\text{IRS}}=\sqrt{M_y^2+M_z^2}d$ denotes the aperture of the IRS. In this section, we adopt a general NUSW-based LoS BS-IRS channel model, which is given by \cite{liu2023near}
	\begin{equation}\label{eqn_RecPow}
		\bs{h}_{\text{BI}}(\bs{t})=\left[\frac{\lambda}{4\pi D(\bs{t},m_y,m_z)}e^{j\frac{2\pi}{\lambda}D(\bs{t},m_y,m_z)}\right]_{m_y\ic{M}_y,m_z\ic{M}_z},
	\end{equation}
	where $D(\bs{t},m_y,m_z)=||\bs{t}-\bs{e}_{m_y,m_z}||$ denotes the distance between the single MA and the $(m_y,m_z)$-th IRS reflecting element. Thus, the effective BS-user channel power gain at the user can be expressed as $G_u(\bs{t},\bs{\Phi})=|\bs{h}_{\text{IU}}^H\bs{\Phi}\bs{h}_{\text{BI}}(\bs{t})|^2$. Note that the NUSW-based channel model assumes amplitude variation among the elements of the BS-IRS channel, which is thus more general than its uniform counterpart (assuming no amplitude variation) and the conventional far-field channel model. Hence, the analytical results derived from the NUSW-based model should also hold for the latter two models. The analytical results specific to the far-field channel model will be provided in Section \ref{analysis_far}.
	
	To maximize $G_u(\bs{t},\bs{\Phi})$ for any given $\bs{t}$, the optimal IRS reflection matrix should guarantee that the BS-IRS channel and the IRS-user channel are in-phase, i.e.,
	\begin{equation}\label{eqn_OptIRS_Phase}
		\left[\bs{\Phi}\right]_{m,m}=\arg\left([\bs{h}_{\text{IU}}]_m\right)-\arg\left([\bs{h}_{\text{BI}}(\bs{t})]_m\right), \forall m \ic{M},
	\end{equation}
	where $[\bs{h}_{\text{IU}}]_m$  and $[\bs{h}_{\text{BI}}(\bs{t})]_m$ denote the $m$-th entries of $\bs{h}_{\text{IU}}$ and $\bs{h}_{\text{BI}}(\bs{t})$, respectively. With \eqref{eqn_OptIRS_Phase}, the effective BS-user channel power gain can be expressed as
	\begin{equation}\label{eqn_RecPow_OptPh}
		G_u(\bs{t})=\left(\frac{\lambda}{4\pi }\right)^2\left(\left|\sum_{m_y=1}^{M_y}\sum_{m_z=1}^{M_z}\frac{|h_{\text{IU},m_y,m_z}|}{D(\bs{t},m_y,m_z)}\right|\right)^2,
	\end{equation}
	where $h_{\text{IU},m_y,m_z}$ denotes the channel from the $(m_y,m_z)$-th IRS reflecting element to the user. In the following, we first analyze the optimal antenna position $\bs{t}$ that maximizes \eqref{eqn_RecPow_OptPh} and then analyze the fluctuation of \eqref{eqn_RecPow_OptPh} within $\ca{C}_t$.
	
	\subsubsection{Optimal Antenna Position}
	To simplify \eqref{eqn_RecPow_OptPh}, we introduce the following lemma.
	
	\begin{lemma}
		Let $R(\bs{t})=\sqrt{x_t^2+y_t^2+z_t^2}$ denote the distance between the MA and the origin. If $\frac{y_td}{R^2(\bs{t})},\frac{z_td}{R^2(\bs{t})}\ll1$,\footnote{Note that these conditions usually hold in practice, as $d$ is wavelength-level and should be much smaller than $R(\bs{t})$, especially for high-frequency wireless communication systems (e.g, millimeter wave and Terahertz wave \cite{ning2023beamforming}).} the channel power gain in \eqref{eqn_RecPow_OptPh} can be approximated as
		\begin{equation}\label{eqn_RecPow_Approx}
			G_u(\bs{t})\approx\left(\frac{\lambda}{4\pi }\right)^2H^2(M_y,M_z,\bs{t}),
		\end{equation}
		where
		\begin{equation}\label{eqn_G}
			H(M_y,M_z,\bs{t})=\sum_{m_y=1}^{M_y}\sum_{m_z=1}^{M_z}\frac{|h_{\text{IU},m_y,m_z}|}{\sqrt{R^2(\bs{t})+(m_yd)^2+(m_zd)^2}}.
		\end{equation}
	\end{lemma}
	\begin{IEEEproof}
		The distance between the MA and the $(m_y,m_z)$-th element can be approximated as
		\begin{align}\label{eqn_Dist_approx}
			&D(\bs{t},m_y,m_z)=||\bs{t}-\bs{e}_{m_y,m_z}||\nonumber\\
			&=\sqrt{x_t^2+(y_t-m_yd)^2+(z_t-m_zd)^2}\nonumber\\
			&=\sqrt{R^2(\bs{t})-2(y_tm_yd+z_tm_zd)+(m_yd)^2+(m_zd)^2}\nonumber\\
			&\approx\sqrt{R^2(\bs{t})+(m_yd)^2+(m_zd)^2},
		\end{align}
		where the approximation is due to $\frac{y_td}{R^2(\bs{t})},\frac{z_td}{R^2(\bs{t})}\ll1$. By substituting \eqref{eqn_Dist_approx} into \eqref{eqn_RecPow_OptPh}, we can obtain \eqref{eqn_RecPow_Approx}.
	\end{IEEEproof}
	
	Based on \eqref{eqn_RecPow_Approx} and \eqref{eqn_G}, it is not difficult to see that $G_u(\bs{t})$ monotonically decreases with $R(\bs{t})$. Thus, if the single MA is deployed at
	\begin{equation}\label{eqn_NearestPos}
		\bs{t}^{\star} = \arg\underset{\bs{t}\ic{C}_t}{\min}\,\,R(\bs{t}),
	\end{equation}
	the channel power gain in \eqref{eqn_RecPow_Approx} can always be maximized. Note that the optimal antenna position in \eqref{eqn_NearestPos} is independent of the IRS-user channel, $\bs{h}_{\text{IU}}$.  For example, assuming that the MA array is parallel to the $x$-axis, it can be shown that $\bs{t}^{\star}=\left[-\frac{A}{2}+x_B,y_B,z_B\right]^T$. As the positions of the IRS and BS are generally fixed in practice, the optimal antenna position in \eqref{eqn_NearestPos} is fixed as well. This indicates that in the case of a single MA, it always yields the maximum SNR at the user by deploying an FPA at \eqref{eqn_NearestPos}, regardless of the IRS-user channel and the BS-IRS distance.\vspace{-1pt}
	
	\begin{figure}[!t]
		\centering
		\captionsetup{justification=raggedright,singlelinecheck=false}
		\subfloat[$N=1$\centering]{
			\includegraphics[width=0.50\linewidth]{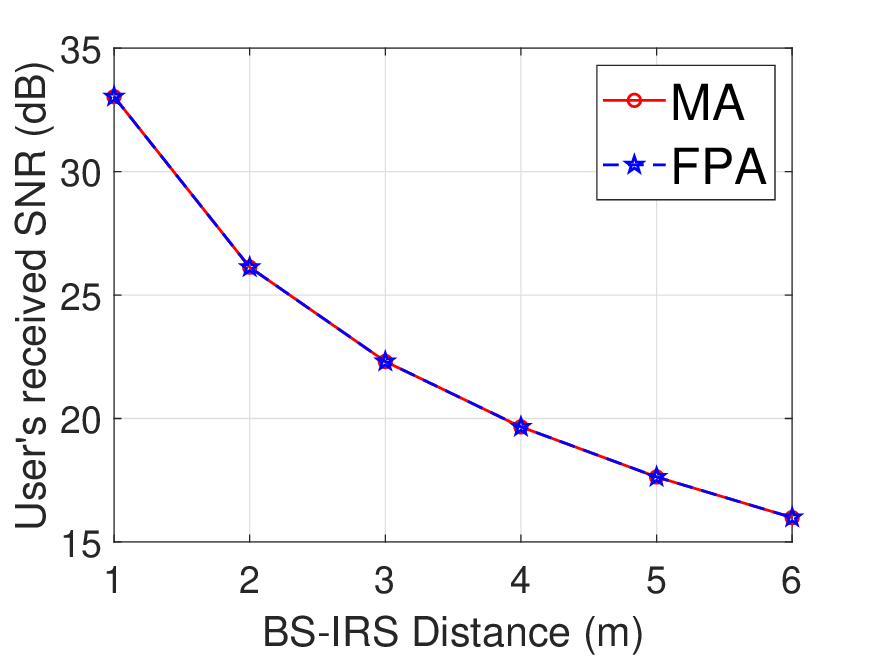}
		}
		\hfil
		\hspace{-20pt}
		\subfloat[$N=4$\centering]{
			\includegraphics[width=0.50\linewidth]{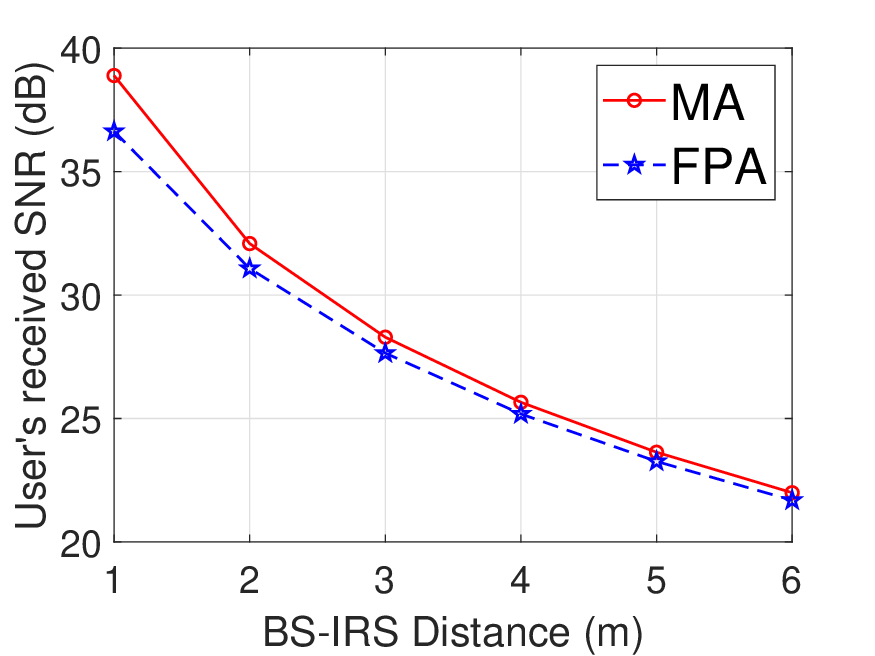}
		}
		\captionsetup{font=footnotesize}
		\caption{Received SNRs at the user in the LoS BS-IRS channel.}
		\vspace{-20pt}
		\label{Fig_LoS_SU}
	\end{figure}
	
	To verify the above analyses, we plot in Fig. 2(a) the received SNR at the user versus the BS-IRS distance under the NUSW-based LoS BS-IRS channel. The BS is equipped with a single MA. The operating frequency is $f=5$ GHz. The total number of the IRS reflecting elements is $M=25^2$. The transmit SNR is $P/\sigma^2=110$ dB. In the FPA benchmark scheme, the antenna is fixed at \eqref{eqn_NearestPos}. It is observed that the SNR achievable by a single MA and a single FPA are identical for all BS-IRS distances considered, which validates our analyses.
	
	\begin{remark}
		Different from the above case with an IRS, if we consider an NUSW-based direct LoS channel from the single-MA BS to the user, the BS-user channel is given by
		\begin{equation}\label{eqn_Ch_BU}
			h_{\text{BU}}(\bs{t})=\frac{\lambda}{4\pi||\bs{t}-\bs{q}_U||_2}e^{j\frac{2\pi}{\lambda}||\bs{t}-\bs{q}_U||_2},
		\end{equation}
		where $\bs{q}_U=[x_U,y_U,z_U]^T\ib{R}^{3\times1}$ denotes the coordinate of the user. It can be verified that the optimal antenna position that maximizes \eqref{eqn_Ch_BU} is given by
		\begin{equation}
			\bs{t}^{\star} = \arg\underset{\bs{t}\ic{C}_t}{\min}\,\,||\bs{t}-\bs{q}_U||_2.
		\end{equation}
		In the case that $\ca{C}_t$ is parallel to the $x$-axis, we have
		\begin{equation}
			\boldsymbol{t}^{\star}=\begin{cases}
				\left[ -\frac{A}{2}+x_B,y_B,z_B \right] ^T,\text{if}\,\, x_U\le -\frac{A}{2},\\
				\left[ x_U,y_B,z_B \right] ^T,\quad\qquad\text{if}\,\, -\frac{A}{2}<x_U\le \frac{A}{2},\\
				\left[ \frac{A}{2}+x_B,y_B,z_B \right] ^T,\,\,\,\,\text{otherwise}\\
			\end{cases}
		\end{equation}
		which depends on the user's coordinate $\bs{q}_U$ (or $x_U$), unlike the IRS-aided case.
	\end{remark}
	
	\subsubsection{Fluctuation of Channel Power Gain}
	In this subsection, we analyze the fluctuation of \eqref{eqn_RecPow_Approx} within $\ca{C}_t$. Notably, a more significant variation of the channel power gain generally results in a more pronounced spatial diversity within $\ca{C}_t$ (and hence the performance gain of MAs over FPAs in general). To characterize the fluctuation, we derive the difference of \eqref{eqn_RecPow_Approx} with respect to (w.r.t.) two arbitrary locations in $\ca{C}_t$, denoted as $\bs{t}_1$ and $\bs{t}_2$ ($\bs{t}_1\ne\bs{t}_2$). Without loss of generality, we assume that $\bs{t}_1$ is closer to the origin than $\bs{t}_2$, i.e., $R(\bs{t}_1) < R(\bs{t}_2)$. Then, the difference between the maximum channel power gains at these two positions is given by
	\begin{equation}
		\begin{aligned}\label{eqn_Pu_Diff}
			G_u(\bs{t}_1)-G_u(\bs{t}_2)&=H^2(M_y,M_z,\bs{t}_1)-H^2(M_y,M_z,\bs{t}_2)\\
			&=\left[H(M_y,M_z,\bs{t}_1)+H(M_y,M_z,\bs{t}_2)\right]\\
			&\quad\times\left[H(M_y,M_z,\bs{t}_1)-H(M_y,M_z,\bs{t}_2)\right],
		\end{aligned}
	\end{equation}
	where the constant scalar $\left(\frac{\lambda}{4\pi }\right)^2$ is omitted for simplicity.
	
	Based on \eqref{eqn_Pu_Diff}, we analyze the effects of different key system parameters on it. First, it is easy to see that $H(M_y,M_z,\bs{t}_1)\pm H(M_y,M_z,\bs{t}_2)$ monotonically increases with $M_y$ and $M_z$, provided that $R(\bs{t}_1) < R(\bs{t}_2)$. Thus, $G_u(\bs{t}_1)-G_u(\bs{t}_2)$ also monotonically increases with $M_y$ and $M_z$, for any given $\bs{t}_1$ and $\bs{t}_2$. This indicates that by increasing the size of the IRS, the maximum channel power gains within $\ca{C}_t$ experience more significant fluctuation.
	
	On the other hand, we analyze the effects of the BS-IRS distance on \eqref{eqn_Pu_Diff}. For simplicity, we assume that the MA array is perpendicular to the IRS. Then, $R(\bs{t}_i)$ can be expressed as $R(\bs{t}_i)=\sqrt{d_{\text{BI}}^2+d^2(\bs{t}_i)}$, $i=1,2$, where $d(\bs{t}_i)=||\bs{q}_B-\bs{t}_i||_2$ denotes the distance between the $i$-th location and the center of the MA array, and $d_{\text{BI}}$ denotes the distance from the center of $\ca{C}_t$ to that of the IRS. As $R(\bs{t}_1)<R(\bs{t}_2)$, $H(M_y,M_z,\bs{t}_1)\pm H(M_y,M_z,\bs{t}_2)$ monotonically decreases with $d_{\text{BI}}$. Thus, $G_u(\bs{t}_1)-G_u(\bs{t}_2)$ also decreases with $d_{\text{BI}}$. This indicates that a more/less significant fluctuation of \eqref{eqn_RecPow_Approx} will be resulted if the BS-IRS distance reduces/increases.
	
	Based on the above, although the antenna position yielding the maximum channel power gain is fixed as \eqref{eqn_NearestPos}, the channel power gain within $\ca{C}_t$ may fluctuate with different degrees. As such, it can be inferred that in the case of multiple antennas, the performance gain of MAs over FPAs may still exist, unlike the single-MA case. To verify this claim, we plot in Fig. 2(b) the received SNR at the user versus the BS-IRS distance with $N=4$ MAs and FPAs, with other simulation parameters the same as those in Fig. 2(a). In the FPA benchmark, the $N$ antennas are arranged in a uniform linear array and separated by half-wavelength. The IRS reflection (for both MAs and FPAs) and antenna positions (for MAs) are optimized based on the algorithms to be presented in Section III-B. It is observed from Fig. 2(b) that unlike the single-MA case, employing multiple MAs can still yield a performance gain over FPAs, especially if the BS-IRS distance is small. Moreover, the performance gain is observed to decrease with the BS-IRS distance, which is consistent with our previous analyses.
	
	\subsubsection{Far-Field BS-IRS Channel}\label{analysis_far}
	All of the above analytical results are derived under the general NUSW-based BS-IRS channel model. Next, we consider a special case with the far-field BS-IRS channel. In this case, the movement region is approximated as a point source; thus, the distance between the MA and the $(m_y,m_z)$-th IRS reflecting element in \eqref{eqn_Dist_Approx} can be approximated as identical, i.e.,
	\begin{equation}\label{eqn_Dist_Approx}
		\begin{aligned}
			D(\bs{t},m_y,m_z)&\approx\sqrt{R^2(\bs{t})+(m_yd)^2+(m_zd)^2}\\
			&\approx d_{\text{BI}},\,\,\forall m_y,m_z.
		\end{aligned}
	\end{equation}
	By substituting \eqref{eqn_Dist_Approx} into \eqref{eqn_Pu_Diff}, it can be seen that $G_u(\bs{t}_1)-G_u(\bs{t}_2)=0$, $\forall \bs{t}_1, \bs{t}_2$. This implies that in the far-field scenario, the channel power gain within $\ca{C}_t$ is uniform. As a result, even employing multiple MAs may not achieve any performance gain over FPAs, as can be rigorously shown below.
	
	Given a multi-MA BS, the BS-IRS channel can be expressed as $\bs{H}_{\text{BI}}(\bs{T})=\beta_{\text{BI}}\bs{u}\bs{v}^H(\bs{T})$, where $\beta_{\text{BI}}$ encompasses the large-scale path gain, $\bs{u}\ib{C}^{M\times1}$ and $\bs{v}(\bs{T})\ib{C}^{N\times1}$ denote the receive and transmit array responses at the IRS and BS, respectively. Accordingly, the channel power gain can be expressed as $G_u(\bs{T},\bs{\Phi})=\left|\beta_{\text{BI}}|^2|\bs{h}_{\text{IU}}\bs{\Phi}\bs{u}\right|^2\left|\bs{v}^H(\bs{T})\bs{w}\right|^2$. It can be seen that for any given $\bs{T}$, to maximize $G_u(\bs{T},\bs{\Phi})$, the optimal BS active beamforming should be set as 
	\begin{equation}
		\bs{w}(\bs{T})=\frac{\bs{v}(\bs{T})}{||\bs{v}(\bs{T})||_2},
	\end{equation}
	leading to $\left|\bs{v}^H(\bs{T})\bs{w}\right|^2=N$, regardless of $\bs{T}$. Hence, employing multiple MAs cannot yield any performance gain over FPAs in the far-field scenario.
	
	It is noteworthy that we have previously shown in Section III-A2 that multiple MAs can achieve a more substantial performance gain over FPAs as the BS-IRS distance decreases and/or the IRS size increases. This in fact renders the BS-IRS channel condition closer to the near field. Consequently, it can be concluded that in the single-user scenario, the performance gain of multiple MAs over multiple FPAs may become increasingly pronounced as the BS-IRS LoS channel becomes more dominated by near-field propagation.\vspace{-8pt}
	
	\subsection{Proposed Solution to (P2)}
	Next, we focus on the general BS-IRS channel and solve (P2). Note that for any given APV $\bs{T}$ and IRS reflection matrix $\bs{\Phi}$, the optimal BS transmit beamforming is given by the maximum transmission ratio (MRT), i.e.,
	\begin{equation}\label{eqn_MRT}
		\bs{w}(\bs{T},\bs{\Phi})=\frac{\left(\bs{h}_{\text{IU}}^H\bs{\Phi}\bs{H}_{\text{BI}}(\bs{T})\right)^H}{||\bs{h}_{\text{IU}}^H\bs{\Phi}\bs{H}_{\text{BI}}(\bs{T})||_2}.
	\end{equation}
	By substituting \eqref{eqn_MRT} into (P2), it can be expressed as
	\begin{subequations}
		\begin{align}
			{\text{(P2)}}\quad &\underset{\bs{\Phi},\bs{T}}{\max}\quad \gamma(\bs{T},\bs{\Phi})=||\bs{h}_{\text{IU}}^H\bs{\Phi}\bs{H}_{\text{BI}}(\bs{T})||^2_2 \nonumber
			\\
			\mathrm{s.t.}\quad & \text{\eqref{eqn_IRS_Phase_Cons}},\,\, \text{\eqref{eqn_MA_Region_Cons}},\,\, \text{\eqref{eqn_MA_Coordinate_Cons}}, \nonumber
		\end{align}
	\end{subequations}
	where the constant $P/\sigma^2$ is omitted. However, (P2) is still a non-convex optimization problem. Next, we propose an AO algorithm to decompose (P2) into two subproblems and solve them accordingly.
	
	\subsubsection{Optimizing $\bs{\Phi}$ for a Given $\bs{T}$}
	First, we optimize the IRS reflection matrix $\bs{\Phi}$ for any given APV $\bs{T}$. Let $\bs{G}_1\triangleq\text{diag}(\bs{h}_\text{IU}^H)\bs{H}_{\text{BI}}(\bs{T})\ib{C}^{M\times N}$ and $\bs{g}_{1,m}\ib{C}^{1\times N}$ denote the $m$-th ($m\ic{M}$) row of $\bs{G}_1$, which are fixed with a given APV $\bs{T}$. Then, (P2) can be simplified as
	\begin{subequations}
		\begin{align}
			{\text{(P2-1)}}\quad &\underset{\bs{\varphi}}{\max}\quad \gamma(\bs{T},\bs{\Phi})=\left|\left|\sum_{m=1}^{M}{\bs{g}_{1,m}e^{j\varphi_m}}\right|\right|^2_2 \nonumber
			\\
			\mathrm{s.t.}\quad & \varphi_m\in[0,2\pi],\,\,\forall m \ic{M},
		\end{align}
	\end{subequations}
	which, however, is still a non-convex optimization problem. To tackle this non-convexity, we first define $\bs{\alpha}_m=\sum_{i\ne m}^{M}{\bs{g}_{1,i}e^{j\varphi_i}}$, $\forall m\ic{M}$, and rewrite the objective function in (P2-1) as a more tractable form as
	\begin{equation}\label{eqn_P2-1_ObjFun}
		\begin{aligned}
			\gamma(\bs{T},\bs{\Phi})&=\left|\left|\bs{g}_{1,m}e^{j\varphi_m}+\bs{\alpha}_m\right|\right|^2_2\\
			&=\left(\bs{g}_{1,m}e^{j\varphi_m}+\bs{\alpha}_m\right)\left(\bs{g}_{1,m}e^{j\varphi_m}+\bs{\alpha}_m\right)^H\\
			&=\left|\left|\bs{g}_{1,m}\right|\right|^2_2+\left|\left|\bs{\alpha}_m\right|\right|^2_2+2\Re\left\{\bs{\alpha}_m\bs{g}_{1,m}^He^{-j\varphi_m}\right\}.
		\end{aligned}
	\end{equation}	
	Next, we propose to utilize an element-wise BCD method to optimize the IRS reflecting coefficients sequentially. Specifically, it can be easily shown from \eqref{eqn_P2-1_ObjFun} that for any given $\varphi_i$, $i \in {\cal M}$, $i \ne m$, the optimal $\varphi_m$ for (P2-1) should be such that the phase of $\bs{\alpha}_m\bs{g}_{1,m}^H$ and $e^{-j\varphi_m}$ are identical, which is given by
	\begin{equation}\label{eqn_OptPhase}
		\varphi_m^{\star}=\arg\left(\bs{\alpha}_m\bs{g}_{1,m}^H\right).
	\end{equation}
	As such, in the $m$-th BCD iteration, we can fix $\varphi_i$, $i \ne m$, $i \in {\cal M}$ and optimize $\varphi_m$ as \eqref{eqn_OptPhase}. Then, the BCD iteration proceeds until the phase shifts of all $M$ IRS reflecting elements have been updated.
	
	\subsubsection{Optimizing $\bs{T}$ for a Given $\bs{\Phi}$}
	Next, we optimize the APV $\bs{T}$ with a given IRS reflection matrix $\bs{\Phi}$. Let $\bs{h}_{\text{BI}}(\bs{t}_n)\ib{C}^{M\times1}$ denote the $n$-th column of $\bs{H}_{\text{BI}}(\bs{T})$. Then, the objective function of (P2), i.e., \eqref{eqn_SNR}, can be recast as
	\begin{equation}
		\begin{aligned}
			\gamma(\bs{T},\bs{\Phi})=\sum_{n=1}^{N}{\left|\bs{g}_2^H\bs{h}_{\text{BI}}(\bs{t}_n)\right|^2},
		\end{aligned}
	\end{equation}
	where $\bs{g}_2^H=\bs{h}_{\text{IU}}^H\bs{\Phi}$. As such, (P1) can be simplified as
	\begin{subequations}\label{eqn_OptPrblm_P3}
		\begin{align}
			{\text{(P2-2)}}\quad &\underset{\bs{T}}{\max}\,\, \sum_{n=1}^{N}{\left|\bs{g}_2^H\bs{h}_{\text{BI}}(\bs{t}_n)\right|^2} \quad\mathrm{s.t.}\,\,\,\, \text{\eqref{eqn_MA_Region_Cons}},\,\,\text{\eqref{eqn_MA_Coordinate_Cons}},\nonumber
		\end{align}
	\end{subequations}
	which can be solved \emph{optimally} by applying the graph-based approach proposed in our previous work \cite{mei2024movable} by modifying the weight assignment therein.
	
	Specifically, we uniformly sample the transmit array into $L_{\text{samp}}$ discrete sampling points with an equal spacing $\delta_s=A/L_{\text{samp}}$ between any two adjacent sampling points. Let $\bs{p}_l$, $\bs{p}_l\ib{R}^{3\times1}$ denote the coordinate of the $l$-th sampling point, $l\ic{L}\triangleq\{1,2,\cdots,L_\text{samp}\}$, and $a_n$ denote the index of the selected sampling point for the $n$-th MA, $a_n\ic{L}$. Then, we can transform (P2-2) into the following discrete sampling point selection problem,
	\begin{subequations}\label{eqn_OptPrblm_P4}
		\begin{align}
			{\text{(P2-3)}}\quad &\underset{\left\{a_n\right\}}{\max}\,\, \sum_{n=1}^{N}{\left|g_{a_n}\right|^2}\nonumber\\
			&\,\,\mathrm{s.t.}\,\,\,\, a_n\ic{L},\\
			&\,\,\quad\quad|a_i-a_j|>a_{\min}, \forall i,j\ic{N}, i\ne j,
		\end{align}
	\end{subequations}
	where $g_{a_n}=\bs{g}_2^H\bs{h}_{\text{BI}}(\bs{p}_{a_n})$, $\forall n\ic{N}$, and $a_{\min}=D_{\min}/\delta_s\gg1$. Next, we can construct a directed weighted graph to equivalently transform (P2-3) into a fixed-hop shortest path problem that can be optimally solved in polynomial time using the dynamic programming. The details are omitted for brevity, for which interested readers can refer to \cite{mei2024movable}. It is worth noting that the graph-based algorithm is general and applicable to any type of channel model for MAs.
	
	Based on the above, we can alternately solve (P2-1) and (P2-3) by applying the element-wise BCD method and the graph-based approach. The overall algorithm for solving (P2) is summarized in Algorithm~\ref{alg_SU}. As this process always yields a non-decreasing objective value of (P2), the convergence of AO is guaranteed. Next, we analyze the computational complexity of Algorithm 1. The computational complexity of optimizing $\bs{\Phi}$ is given by $\ca{O}(MI_1)$ with $I_1$ denoting the number of BCD iterations, whereas that of optimizing $\bs{T}$ is given by $\ca{O}(NL_{\text{samp}}^2)$. Thus, the overall complexity of Algorithm 1 is given by $\ca{O}(MI_1+NL_{\text{samp}}^2)$.\vspace{-6pt}
	\begin{algorithm}[!t]
		\caption{Proposed AO Algorithm for Solving (P2)}
		\label{alg_SU}
		\begin{algorithmic}[1]
			\STATE Initialize $\bs{\Phi}^{(0)}$, $\bs{T}^{(0)}$, and convergence accuracy $\epsilon_1$ and $\epsilon_2$.
			\STATE Set $l=0$.
			\REPEAT
				\STATE $\bs{\Phi}^{(l+1)}\leftarrow\bs{\Phi}^{(l)}$.
				\REPEAT
					\FOR{$m=1\rightarrow M$}
						\STATE Calculate $\varphi_m^{\star}$ according to \eqref{eqn_OptPhase} with $\bs{\Phi}^{(l+1)}$.
						\STATE $\left[\bs{\Phi}^{(l+1)}\right]_{m,m}\leftarrow\varphi_m^{\star}$.
					\ENDFOR
				\UNTIL{the increment of the objective value of (P2-1) is below the given convergence accuracy $\epsilon_1$.}
				\STATE Calculate the received SNR $\gamma_1^{(l+1)}$ with $\bs{\Phi}^{(l+1)}$ and $\bs{T}^{(l)}$ according to \eqref{eqn_SNR}.
				\STATE Obtain $\bs{T}^{(l+1)}$ by solving (P2-3) via the graph-based method.
				\STATE Calculate the received SNR $\gamma_2^{(l+1)}$ with $\bs{\Phi}^{(l+1)}$ and $\bs{T}^{(l+1)}$ according to \eqref{eqn_SNR}.
				\STATE Set $l=l+1$.
			\UNTIL{$\left|\gamma_2^{(l)}-\gamma_1^{(l)}\right|\le\epsilon_2$.}
			\STATE Output $\bs{\Phi}^{(l)}$, $\bs{T}^{(l)}$ as the optimized solutions to (P2).
		\end{algorithmic}
	\end{algorithm}
	
	\section{Multi-User Case}
	In this section, we conduct the performance analysis of the general multi-user case and solve (P1) accordingly.\vspace{-9pt}
	
	\subsection{Performance Analysis}
	To reveal useful insights, we first assume a far-field LoS BS-IRS channel as in Section III-A3 and characterize the performance gain of MAs over FPAs in an IRS-aided multi-user system. In this case, we have $\bs{H}_{\text{BI}}=\beta_{\text{BI}}\bs{u}\bs{v}^H(\bs{T})$ and the cascaded channel from the BS to user $k$ can be expressed as $\bs{h}_k^H(\bs{T})=\bs{h}_{\text{IU}}^H\bs{\Phi}\bs{H}_{\text{BI}}(\bs{T})\triangleq q_k\bs{v}^H(\bs{T})$, where $q_k=\beta_{\text{BI}}\bs{h}_{\text{IU}}^H\bs{\Phi}\bs{u}\ib{C}$, $k\ic{K}$.
	
	Unlike the single-user case, it is generally difficult to acquire the optimal BS transmit precoding in closed-form in the multi-user case even for any given IRS reflection. As such, we assume a general regularized ZF (RZF) precoding at the BS \cite{yang2024flexible}. Specifically, by stacking the cascaded channel vectors of all users in rows, we define $\bs{H}=\left[\bs{h}_1(\bs{T}),\bs{h}_2(\bs{T}),\cdots,\bs{h}_K(\bs{T})\right]^H=\bs{q}\bs{v}^H(\bs{T})\ib{C}^{K\times N}$, where $\bs{q}=[q_1,q_2,\cdots,q_K]^T\ib{C}^{K\times1}$. Then, the RZF precoding is given by
	\begin{equation}\label{eqn_RZF_PreCoding}
		\bs{W}^{\text{RZF}}=\bs{H}^H\left(\bs{H}\bs{H}^H+\alpha\bs{I}\right)^{-1},
	\end{equation}
	where $\alpha$ is the regularization factor. Note that $\bs{W}^{\text{RZF}}$ reduces to the MRT, ZF, and MMSE precoding by setting specific values of $\alpha$, i.e.,
	\begin{equation}
		\bs{W}^{\text{RZF}}=\begin{cases}
			\bs{W}^{\text{MRT}}\triangleq\bs{H}^H,\qquad\qquad\qquad\qquad\,\,\,\,\,\alpha\rightarrow\infty,\\
			\bs{W}^{\text{ZF}}\triangleq\bs{H}^H\left(\bs{H}\bs{H}^H\right)^{-1},\qquad\qquad\alpha=0,\\
			\bs{W}^{\text{MMSE}}\triangleq\bs{H}^H\left(\bs{H}\bs{H}^H+\sigma^2\bs{I}\right)^{-1},\alpha=\sigma^2.\\
		\end{cases}
	\end{equation}
	However, it is still difficult to analyze the effects of the RZF precoding due to its complicated expressions. To address this issue, we first transform \eqref{eqn_RZF_PreCoding} into a more tractable form and then obtain a simplified expression of the transmit precoding vector for each user.
	
	To this end, we first calculate the matrix inversion in \eqref{eqn_RZF_PreCoding}. Let $\bs{F}=\left(\bs{H}\bs{H}^H+\alpha\bs{I}\right)^{-1}$ and note that
	\begin{equation}\label{eqn_InvMat}
		\bs{F}\!=\!\left(\bs{q}\bs{v}^H(\bs{T})\bs{v}(\bs{T})\bs{q}^H+\alpha\bs{I}\right)^{-1}
			=\left(N\bs{q}\bs{q}^H+\alpha\bs{I}\right)^{-1},
	\end{equation}
	which is independent of the APV $\bs{T}$. By substituting \eqref{eqn_InvMat} into \eqref{eqn_RZF_PreCoding}, we can obtain
	\begin{equation}\label{eqn_RZF_PreCoding_New}
		\bs{W}^{\text{RZF}}=\bs{v}(\bs{T})\bs{q}^H\bs{F}.
	\end{equation}
	Let $\bs{\hat{q}}^H\triangleq\bs{q}^H\bs{F}\ib{C}^{1\times K}$ and $\hat{q}_k$ denote the $k$-th entry of $\bs{\hat{q}}$. As such, the RZF precoder for user $k$ can be expressed as
	\begin{equation}\label{eqn_RZF_PreVec}
		\bs{w}^{\text{RZF}}_k(\bs{T})=\sqrt{p}_k\frac{\hat{q}_k^*\bs{v}(\bs{T})}{\left|\left|\hat{q}_k^*\bs{v}(\bs{T})\right|\right|_2}=\sqrt{\frac{p_k}{N}}e^{-j\angle \hat{q}_k}\bs{v}(\bs{T}),
	\end{equation}
	where $p_k$ denotes the transmit power allocated to user $k$. By substituting \eqref{eqn_RZF_PreVec} into \eqref{eqn_SINR}, the achievable rate of user $k$ with the RZF precoding is given by
	\begin{equation}\label{eqn_AcheRate_FFLoS}
		R_k^{\text{RZF}}=\log_2\left(1+\frac{Np_k\left|q_k\right|^2}{N\left|q_k\right|^2\sum_{i\ne k}{p_i}+\sigma^2}\right), \,\, k\ic{K}.
	\end{equation}
	From \eqref{eqn_AcheRate_FFLoS}, it can be observed that the achievable rate of user $k$ is \emph{irrelevant to} the APV $\bs{T}$. This indicates that in the far-field LoS BS-IRS channel and typical transmit precoding schemes (e.g., RZF, ZF, MMSE, and MRT), the MAs cannot yield any performance gain over the conventional FPAs for any given IRS reflection and transmit power allocation.
	
	\begin{remark}
		The main reason for the absence of the performance gain of MAs lies in the fact that all BS-user channels share the same BS-IRS channel. In contrast, in the case without the IRS, if we consider a far-field LoS channel from the multi-MA BS to all users, the BS-user $k$ channel is expressed as
		\begin{equation}
			\bs{h}_{\text{BU},k}(\bs{T})=\beta_{\text{BU},k}\bs{\hat{v}}_k(\bs{T}),\,\,\forall k\ic{K},
		\end{equation}
		where $\bs{\hat{v}}_k(\bs{T})$ denotes the transmit array response at the BS for user $k$, and $\beta_{\text{BU},k}$ denotes the complex path gain of the BS-user $k$ channel. Take the MRT precoding for each user $k$, $k\ic{K}$, as an example, which is given by
		\begin{equation}
			\bs{w}_k^{\text{MRT}}(\bs{T})=\sqrt{\frac{p_k}{N}}e^{j\angle \beta_{\text{BU},k}}\bs{v}_k(\bs{T}).
		\end{equation}
		In this case, the achievable rate of user $k$ is given by
		\begin{equation}\label{eqn_AcheRate_MU_NoIRS}
			\small
			R_k^{\text{MRT}}(\bs{T})=\log_2\left(1+\frac{p_k|\beta_{\text{BU},k}|^2}{|\beta_{\text{BU},k}|^2\sum_{i\ne k}{p_i\left|\bs{v}_k^H(\bs{T})\bs{v}_i(\bs{T})\right|^2+\sigma^2}}\right).
		\end{equation}
		Notably, \eqref{eqn_AcheRate_MU_NoIRS} depends on the channel correlation $\left|\bs{v}_k^H(\bs{T})\bs{v}_i(\bs{T})\right|^2$, which can be tuned by changing the MA positions. Hence, without the IRS, the MAs can still provide performance gain over the conventional FPAs, unlike the case with the IRS.\vspace{-15pt}
	\end{remark}
	
	\subsection{Proposed Solution to (P1)}\label{Alg_MU}
	Next, we focus on solving (P1). Due to the complicated coupling among the variables, we propose an AO algorithm to decompose it into three subproblems and solve them accordingly.\vspace{1pt}
	
	\subsubsection{Optimizing $\bs{W}$ for Given $\mbf{\Phi}$ and $\bs{T}$}
	First, for any given IRS reflection matrix $\bs{\Phi}$ and APV $\bs{T}$, (P1) becomes equivalent to the conventional weighted sum-rate maximization problem in multi-user MISO systems with an identical weight for each user, which has been extensively investigated in previous works \cite{shi2011iteratively,guo2020weighted}. In this paper, we apply the WMMSE algorithm to obtain a high-quality sub-optimal solution. Specifically, by introducing two auxiliary variables $\bs{\chi}=\left[\chi_1,\chi_2,\cdots,\chi_K\right]^T\ib{C}^{K\times1}$ and $\bs{\kappa}=\left[\kappa_1,\kappa_2,\cdots,\kappa_K\right]^T\ib{C}^{K\times1}$, (P1) can be equivalently transformed into the following optimization problem\vspace{-3pt}
	\begin{subequations}
		\begin{align}
			{\text{(P3-1)}}\quad \underset{\bs{\chi},\bs{\kappa},\bs{W}}{\max}\quad f_1(\bs{\chi},\bs{\kappa},\bs{W}) \nonumber \quad \mathrm{s.t.}\quad \eqref{eqn_Pw_Cons},
		\end{align}
	\end{subequations}
	where $f_1(\bs{\chi},\bs{\kappa},\bs{W})=\sum_{k=1}^{K}\kappa_ku(\chi_k,\bs{W})-\log\kappa_k$ and $u(\chi_k,\bs{W})=|\chi_k|^2(\sum_{i=1}^{K}\left|\bs{h}_k^H\bs{w}_i\right|^2+\sigma^2)-\mathrm{Re}\{\chi_k^*\bs{h}_k^H\bs{w}_k\}+1$. After the above transformation, the original problem becomes more tractable and can be efficiently solved by updating $\bs{\chi}$, $\bs{\kappa}$,  and $\bs{W}$ iteratively. Specifically, in the $(l+1)$-th iteration of the WMMSE algorithm, these three variables are calculated based on the following updating rules:\vspace{-6pt}
	\begin{subequations}\label{eqn_WMMSE}
		\begin{align}
			&\chi_k^{(l+1)}=\left(\sum_{i=1}^{K}\left|\bs{h}_k^H(\bs{T})\bs{w}_i^{(l)}\right|^2+\sigma^2\right)^{-1}\bs{h}_k^H(\bs{T})\bs{w}_k^{(l)},\\
			&\kappa_k^{(l+1)}=\left(1-\chi_k^{*(l+1)}\bs{h}_k^H(\bs{T})\bs{w}_k^{(l)}\right)^{-1},\\
			&\bs{w}_k^{(l+1)}=\chi_k^{(l+1)}\kappa_k^{(l+1)}\Big(\mu\bs{I}_N+\sum_{i=1}^{K}\left|\chi_k^{(l+1)}\right|^2\kappa_k^{(l+1)}\nonumber\\
			&\qquad\qquad\qquad\times\bs{h}_k(\bs{T})\bs{h}_k^H(\bs{T})\Big)^{-1}\bs{h}_k^H(\bs{T}),
		\end{align}
	\end{subequations}
	where $\mu\ge0$ is the optimal dual variable that ensures $\sum_{k=1}^{K}\left\|\bs{w}_k\right\|_2^2\le P$. Note that $\sum_{k=1}^{K}\left\|\bs{w}_k\right\|_2^2$ is a monotonically decreasing function w.r.t. $\mu$ \cite{sun2024optimization}. Thus, we can find $\mu$ via a bisection search.
	
	\subsubsection{Optimizing $\bs{\Phi}$ for Given $\bs{W}$ and $\bs{T}$}
	First, we define $\bs{r}_{k,i}=\text{diag}(\bs{h}_{\text{IU},k}^H)\bs{H}_{\text{BI}}\bs{w}_i\ib{C}^{M\times1}$, $\forall i,k\ic{K}$. Then, for any given transmit precoding matrix $\bs{W}$ and APV $\bs{T}$, (P1) degenerates into the following optimization problem:\vspace{-3pt}
	\begin{subequations}
		\begin{align}
			{\text{(P3-2)}}\,\, &\underset{\bs{\varphi}}{\min}\,\, f_2(\bs{\varphi})=-\sum_{k\ic{K}}\log_2\Big(1+\frac{|\bs{\varphi}^H\bs{r}_{k,k}|^2}{\sum_{i\ne k}{|\bs{\varphi}^H\bs{r}_{k,i}|^2+\sigma^2}}\Big), \nonumber
			\\
			\mathrm{s.t.}\quad& \eqref{eqn_IRS_Phase_Cons},\nonumber
		\end{align}
	\end{subequations}
	which is still challenging to be solved optimally due to the non-convex objective function and unit-modulus constraints in \eqref{eqn_IRS_Phase_Cons}. Fortunately, a variety of optimization techniques have been proposed to solve such an IRS passive beamforming optimization problem, such as semidefinite relaxation (SDR) \cite{wu2019intelligent,wang2020intelligent}. However, it only yields an approximate solution without optimality guarantee. To strike a balance between the performance and convergence, we apply the CG-based manifold approach \cite{yu2020robust} to solve (P3-2). Specifically, the search space of (P3-2) is a complex circle manifold given by
	\begin{equation}
		\ca{S}^M\triangleq\left\{\mbf{x}\ib{C}^M:\left|x_l\right|=1,l=1,2,\cdots,M\right\},
	\end{equation}
	where $x_l$ is the $l$-th element of $\mbf{x}$. Next, three key steps are required in each iteration of the CG-based manifold approach, as detailed below.
	
	First, we calculate the Riemannian gradient of $f_2(\bs{\varphi})$ denoted as $\mathrm{grad}f_2(\bs{\varphi})$, which is the orthogonal projection of the Euclidean gradient, i.e., $\nabla f_2(\bs{\varphi})$, onto the tangent space of $\ca{S}^M$ at $\bs{\varphi}$. It can be shown that $\nabla f_2(\bs{\varphi}_l)=\sum_{k\ic{K}}{\bs{A}_k(\bs{\varphi}_l)}$, where
	\begin{equation}
		\bs{A}_k(\bs{\varphi})=\frac{\sum_{i\ic{K}}{\bs{r}_{k,i}\bs{r}_{k,i}^H\bs{\varphi}}}{\sum_{i\ic{K}}{\left|\bs{\varphi}^H\bs{r}_{k,i}\right|^2+\sigma^2}}-\frac{\sum_{i\ne k}{\bs{r}_{k,i}\bs{r}_{k,i}^H\bs{\varphi}}}{\sum_{i\ne k}{\left|\bs{\varphi}^H\bs{r}_{k,i}\right|^2+\sigma^2}}.
	\end{equation}
	Let
	\begin{equation}
		T_{\bs{\varphi}_l}\ca{S}^M\triangleq\left\{\mbf{z}\ib{C}^M:\Re\left\{\mbf{z}\odot\bs{\varphi}_l\right\}=\bs{0}_M\right\}
	\end{equation}
	denote the tangent space of $\ca{S}^M$ at $\bs{\varphi}_l$. Then, the Riemannian gradient at $\bs{\varphi}_l$ is given by
	\begin{equation}\label{eqn_RCG_S1}
		\mathrm{grad}f_2(\bs{\varphi}_l)=\nabla f_2(\bs{\varphi}_l)-\Re\left\{\nabla f_2(\bs{\varphi}_l)\odot\bs{\varphi}_l^*\right\}\odot\bs{\varphi}_l.
	\end{equation}
	
	Secondly, we can follow the updating rule of the CG method in the Euclidean space to calculate the new search direction, which is given by
	\begin{equation}\label{eqn_CG_Euclid}
		\bs{\eta}_{l+1}=-\nabla f_2(\bs{\varphi}_l)+\tau_l\bs{\eta}_{l},
	\end{equation}
	where $\bs{\eta}_{l}$ denotes the search direction at $\varphi_l$, and $\tau_l$ is the Polak-Ribiere parameter \cite{absil2008optimization}. However, $\bs{\eta}_{l}$ and $\bs{\eta}_{l+1}$ in \eqref{eqn_CG_Euclid} lie in two different tangent spaces $T_{\bs{\varphi}_l}\ca{S}^M$ and $T_{\bs{\varphi}_{l+1}}\ca{S}^M$. Thus, \eqref{eqn_CG_Euclid} cannot be directly applied. To address this issue, we introduce a vector transport function denoted by $\ca{T}(\cdot)$, which is the mapping of a tangent vector from one tangent space to another tangent space, i.e.,
	\begin{equation}
			\ca{T}(\bs{\eta}_l)\triangleq\bs{\eta}_{l}-\Re\left\{\bs{\eta}_l\odot\bs{\varphi}_{l+1}^*\right\}\odot\bs{\varphi}_{l+1}.
	\end{equation}
	Therefore, the updating rule in \eqref{eqn_CG_Euclid} becomes
	\begin{equation}\label{eqn_RCG_S2}
		\bs{\eta}_{l+1}=-\mathrm{grad}f_2(\bs{\varphi}_{l+1})+\tau_l\ca{T}(\bs{\eta}_l).
	\end{equation}
	
	Thirdly, after determining the search direction $\bs{\eta}_l$ at $\bs{\varphi}$, we can find the destination on $\ca{S}^M$ via a retraction operator, which is a mapping from the tangent space to the manifold itself, i.e.,
	\begin{equation}\label{eqn_RCG_S3}
		\ca{R}(\alpha_l\bs{\eta}_{l})\triangleq \mathrm{unt}\left(\alpha_l\bs{\eta}_{l}\right).
	\end{equation}
	where $\alpha_l$ is the Armijo step size. Based on above procedures, the overall algorithm to solve (P3-2) is summarized in Algorithm 2.
	\begin{algorithm}[!t]
		\caption{CG-Based Manifold Optimization for Solving (P3-2)}
		\label{alg_Phi}
		\begin{algorithmic}[1]
			\STATE Initialize $\bs{\varphi}_0$ and convergence accuracy $\epsilon_2$. Set $l=0$.
			\STATE Calculate $\bs{\eta}_{l}=-\mathrm{grad}f_2(\bs{\varphi}_{l})$ according to \eqref{eqn_RCG_S2}.
			\REPEAT
				\STATE Calculate $\bs{\varphi}_{l+1}$ via \eqref{eqn_RCG_S3}.
				\STATE Calculate Riemannian gradient $\mathrm{grad}f_2(\bs{\varphi}_{l+1})$ according to \eqref{eqn_RCG_S1}.
				\STATE Calculate conjugate search direction $\bs{\eta}_{l+1}$ according to \eqref{eqn_RCG_S2}.
				\STATE Set $l=l+1$.
			\UNTIL{$||\mathrm{grad}f_2(\bs{\varphi}_{l})||_2\le\epsilon_2$.}
			\STATE Output $\bs{\varphi}_l$ and set $\bs{\Phi}=\mathrm{diag}\left(\bs{\varphi}_l\right)$.
		\end{algorithmic}
	\end{algorithm}
	
	\subsubsection{Optimizing the APV $\bs{T}$}
	For any given transmit precoding matrix $\bs{W}$ and IRS reflection matrix $\bs{\Phi}$, (P1) is simplified as into the following problem:
	\begin{subequations}
		\begin{align}
			{\text{(P3-3)}}\quad &\underset{\bs{T}}{\max}\quad f_3(\bs{T})=\sum_{k\ic{K}}R_k(\bs{T}),\quad
			\mathrm{s.t.}\quad \eqref{eqn_MA_Region_Cons},\,\,\eqref{eqn_MA_Coordinate_Cons}.\nonumber
		\end{align}
	\end{subequations}
	It can be observed that the objective function $f_3(\bs{T})$ and constraints in \eqref{eqn_MA_Coordinate_Cons} are highly non-linear w.r.t. the APV. To tackle this difficulty, we apply the discrete sampling-based method with sequential search proposed in our previous works \cite{mei2024movable,wei2024joint}, which is applicable to any channel model for MAs and dispenses with the complex gradient computation. In addition, it can yield a comparable performance to the widely applied PSO method in MA position optimization with much lower complexity \cite{wei2024joint}.
	
	Similar to Section IV-B, the transmit array $\ca{C}_t$ is uniformly sampled into $L_{\text{samp}}$ points, with $\bs{p}_l$ ($l\ic{L}$) denoting the coordinate of the $l$-th sampling point. Let $\ca{P}=\left\{\bs{p}_l|l\ic{L}\right\}$ denote the set of all sampling points. Next, we construct a set of initial sampling points denoted by $\hat{\bs{t}}_n$, $n\ic{N}$. In the $n$-th iteration of the sequential search, we only update the coordinate of the $n$-th MA and keep the coordinates of the other $(N-1)$ MAs fixed. Let $\ca{T}_n$ denotes the set of all feasible sampling points in the $n$-th iteration, which is given by
	\begin{equation}\label{eqn_DS_T}
		\ca{P}_n\!\triangleq\!\left\{\left.\bs{p}\ic{P}\right|||\bs{p}-\bs{\hat{t}}_m||_2\ge D_{\min},\forall m\ic{N},m\ne n\right\}.
	\end{equation}
	Let $\hat{\bs{T}}_n=\left[\hat{\bs{t}}_1,\cdots,\bs{t}_n,\cdots,\hat{\bs{t}}_N\right]^T$ denote the collection of the $N$ MAs in the $n$-th iteration. Then, we can optimize $\bs{t}_n$ by solving the following problem:
	\begin{subequations}\label{eqn_DS_Prob}
		\begin{align}
			{\text{(P3-3-}n)}\quad &\underset{\bs{t}_n}{\max}\quad f_3(\hat{\bs{T}}_n),\quad
			\mathrm{s.t.}\quad \bs{t}_n\ic{P}_n,\nonumber
		\end{align}
	\end{subequations}
	which can be optimally solved via an enumeration over $\ca{P}_n$. Let $\hat{\bs{t}}_n^{\star}$ denote the optimal solution to (P3-3-$n$). Next, we can update $\hat{\bs{t}}_n$ as $\hat{\bs{t}}_n^{\star}$ and proceed to solve (P3-3-$(n+1)$). The overall algorithm to solve (P3-3) is summarized in Algorithm 3.
	
	\begin{algorithm}[!t]
		\caption{Discrete Sampling Based Algorithm for Solving (P3-3)}
		\label{alg_APV}
		\begin{algorithmic}[1]
			\STATE Initialize $\hat{\bs{t}}_n$, $n\ic{N}$.
			\FOR{$n=1\rightarrow N$}
				\STATE Construct $\ca{P}_n$ via \eqref{eqn_DS_T}.
				\STATE Obtain $\hat{\bs{t}}_n^{\star}$ by solving (P3-2-$n$).
				\STATE Set $\hat{\bs{t}}_n=\hat{\bs{t}}_n^{\star}$ and $n=n+1$.
			\ENDFOR
			\STATE Output $\hat{\bs{t}}_n$, $n\ic{N}$ as the optimized MAs' positions.
		\end{algorithmic}
	\end{algorithm}
	
	\subsubsection{Overall Algorithm}
	Based on the above, we can execute the proposed AO algorithm to solve (P1). The overall algorithm is summarized in Algorithm 4. To analyze the computational complexity of Algorithm 4, note that the complexity of the WMMSE algorithm is given by $\ca{O}(I_{w}I_{\mu}KN^3)$, where $I_w$ and $I_{\mu}$ denote the iteration numbers of searching $\mu$ and the three-step loop in \eqref{eqn_WMMSE}, respectively \cite{guo2020weighted}. Moreover, the complexity of the CG-based manifold method mainly depends on the calculation of the Euclidean gradient $\bs{\eta}^{(l)}$, and its complexity order is given by $\ca{O}(K^2M^2)$ \cite{guo2020weighted}. In addition, the complexity of the discrete sampling-based method is $\ca{O}(NL_{\text{samp}})$. As a result, the total complexity of the proposed AO algorithm is given by $\ca{O}(I_o(I_{w}I_{\mu}KN^3+I_cK^2M^2+NL_{\text{samp}}))$, where $I_o$ and $I_c$ denote the iteration numbers of the outer loop and the CG-based manifold algorithm, respectively. Moreover, note that the proposed algorithms for all three subproblems of (P1) always yield a non-decreasing objective value of (P1). Hence, the convergence of Algorithm 4 can be guaranteed.\vspace{-9pt}
	\begin{algorithm}[!t]
		\caption{Proposed AO Algorithm for Solving (P1)}
		\label{alg_ALL}
		\begin{algorithmic}[1]
			\STATE Initialize $\bs{W}^{(0)}$, $\bs{\Phi}^{(0)}$, $\bs{T}^{(0)}$ and convergence accuracy $\epsilon_1$. Set $l=0$.
			\STATE Calculate $R_{\mathrm{sum}}^{(i)}=\sum_{k\ic{K}}{R_k(\bs{W}^{(i)},\bs{\Phi}^{(i)},\bs{T}^{(i)})}$.
			\REPEAT
				\STATE Set $i=i+1$.
				\STATE Obtain $\bs{W}^{(i)}$ with given $\bs{\Phi}^{(i-1)}$ and $\bs{T}^{(i-1)}$ according to \eqref{eqn_WMMSE}.
				\STATE Obtain $\bs{\Phi}^{(i)}$ with given $\bs{W}^{(i)}$ and $\bs{T}^{(i-1)}$ according to Algorithm~\ref{alg_Phi}.
				\STATE Obtain $\bs{T}^{(i)}$ with given $\bs{W}^{(i)}$ and $\bs{\Phi}^{(i)}$ according to Algorithm~\ref{alg_APV}.
				\STATE Calculate $R_{\mathrm{sum}}^{(i)}=\sum_{k\ic{K}}{R_k(\bs{W}^{(i)},\bs{\Phi}^{(i)},\bs{T}^{(i)})}$.
			\UNTIL{$\left|R_{\mathrm{sum}}^{(i)}-R_{\mathrm{sum}}^{(i-1)}\right|\le\epsilon_1$.}
			\STATE Output $\bs{W}^{(i)}$, $\bs{\Phi}^{(i)}$ and $\bs{T}^{(i)}$ as the solution to (P1).
		\end{algorithmic}
	\end{algorithm}

	\section{Numerical Results}
	In this section, we provide numerical results to validate our performance analysis and evaluate the efficacy of our proposed algorithms. Unless otherwise specified, the simulation parameters are set as follows. The operating frequency is $f_c=5$ GHz. The number of MAs at the BS is $N=4$. The total number of the IRS reflecting elements along the $y$- and $z$-axes is set identical as $M_y=M_y=15$, leading to $M=M_y\times M_z = 225$. The spacing between any two adjacent IRS reflecting elements is $d=\frac{\lambda}{2}$, and the minimum spacing between any two MAs at the BS is $D_{\min}=\frac{\lambda}{2}$. The length of the transmit region $\ca{C}_t$ is $A=0.6$ m. The spacing of adjacent sampling points in Algorithm~\ref{alg_APV} is $\delta_s=\frac{\lambda}{10}$. The BS's maximum transmit power is $P=46$ dBm, and the average noise power is $\sigma^2=-80$ dBm. The coordinate of the center of the transmit region $\ca{C}_t$ is $\bs{q}_B=[4\sqrt{2},4\sqrt{2},0]^T$. The number of users is set to $K=3$ in the multi-user case. We consider Rician-fading IRS-user $k$ channel, which is given by\vspace{-6pt}
	\begin{equation}
		\bs{h}_{\text{IU},k}=\frac{\lambda}{4\pi d_k^{-\alpha/2}}\left(\sqrt{\frac{\varsigma}{1+\varsigma}}\bs{h}_{\text{IU},k}^{\text{LoS}}+\sqrt{\frac{1}{1+\varsigma}}\bs{h}_{\text{IU},k}^{\text{NLoS}}\right),
	\end{equation}
	where $\bs{h}_{\text{IU},k}^{\text{LoS}}\ib{C}^{M\times1}$ and $\bs{h}_{\text{IU},k}^{\text{NLoS}}\ib{C}^{M\times1}$ represent the LoS and Rayleigh fading components of the IRS-user $k$ channel, respectively. The Rician factor is set to $\varsigma=3$ dB. The distance from the IRS to user $k$, i.e., $d_k$, $k\ic{K}$, is assumed to follow the uniform distribution from 30 m to 50 m, and the path loss exponent for the IRS-user channels is $\alpha=2.8$. Under the parameters above, it follows from \eqref{eqn_RayD} that the maximum BS-IRS Rayleigh distance is 50.95 m, which is much larger than the BS-IRS distance in the simulation, i.e., from 1 m to 6 m. Hence, we consider the NUSW-based multi-path BS-IRS channel. Let $L$ denote the number of dominant paths over the BS-IRS link, and $\bs{s}_{\text{BI},p}\ib{R}^{3\times1}$ denote the coordinate of the $p$-th scatterer, $p=1,2,\cdots,L$. Then, the BS-IRS channel can be expressed as  \cite{liu2023near}
	\begin{equation}\label{eqn_Channel_BI_NF}
		\bs{H}_{\text{BI}}(\bs{T})=l_0\underset{\text{LoS component}}{\underbrace{\bs{H}_{\text{BI}}^{\text{LoS}}(\bs{T})}}+\underset{\text{NLoS components}}{\underbrace{\sum_{p=1}^{L}{l_p\beta_{\text{BI},p}\bs{a}^T(\bs{s}_{\text{BI},p})\bs{b}(\bs{T},\bs{s}_{\text{BI},p}})}},
	\end{equation}
	where the LoS component $\bs{H}_{\text{BI}}^{\text{LoS}}(\bs{T})$ is written as
	\begin{equation}
		\small
		\bs{H}_{\text{BI}}^{\text{LoS}}(\bs{T})=\left[\frac{\lambda}{4\pi D(\bs{t}_n,m_y,m_z)}e^{j\frac{2\pi}{\lambda}D(\bs{t}_n,m_y,m_z)}\right]_{m_y\ic{M}_y,m_z\ic{M}_z}^{n\ic{N}},
	\end{equation}
	with $D(\bs{t}_n,m_y,m_z)=||\bs{t}_n-\bs{e}_{m_y,m_z}||$ denoting the distance between the $n$-th MA and the $(m_y,m_z)$-th IRS reflecting element, $\forall n,m_y,m_z$. The parameter $l_0$ ($l_p$) denotes the ratio of the average channel power gain of the LoS path (the $p$-th scattered path) to that of all paths, and we assume $l_p\sim\ca{CN}(0,\frac{1}{L+1})$, $p=0,1,2,\cdots,L$. In addition, the vectors $\bs{a}(\bs{s}_{\text{BI},l})\ib{C}^{1\times M}$ and $\bs{b}(\bs{T},\bs{s}_{\text{BI},l})\ib{C}^{1\times N}$ denote the receive and transmit near-field array responses at the IRS and the BS, respectively, which are given by
	\begin{subequations}
		\begin{align}
			\bs{a}(\bs{s})&=\left[e^{j\frac{2\pi}{\lambda}||\bs{s}-\bs{e}_{m_y,m_z}||}\right]_{m_y\ic{M}_y,m_z\ic{M}_z},\\
			\bs{b}(\bs{T},\bs{s})&=\left[e^{j\frac{2\pi}{\lambda}||\bs{s}-\bs{t}_n||}\right]_{n\ic{N}}.\label{eqn_NF_Vec}
		\end{align}
	\end{subequations}
	
	All the results to be shown are averaged over $10^3$ independent channel realizations. Moreover, we consider the following benchmark schemes:
	\begin{enumerate}
		\item \textbf{FPA}: The $N$ MAs are deployed symmetrically to $\bs{q}_B$ and separated by the minimum distance $D_{\text{min}}$. The transmit precoding matrix $\bs{W}$ and IRS reflection matrix $\bs{\Phi}$ are alternately optimized via a similar process as in Section \ref{Alg_MU}.
		\item \textbf{Antenna selection (AS)}: In this benchmark, $A/D_{\min}$ FPAs are deployed within $\ca{C}_t$ and separated by the minimum distance $D_{\min}$. Among them, $N$ antennas are selected for transmission. The associated optimization problem can be solved by applying a similar AO algorithm as in Section \ref{Alg_MU} by setting $\delta_s=D_{\min}$.
		\item \textbf{MAs with random IRS phase shifts (MA-RPS)}: The IRS reflection matrix is randomly generated as $\left[\bs{\Phi}\right]_{m,m}=e^{j\varphi_m}$, $\varphi_m\sim\ca{U}(0,2\pi)$, $m\ic{M}$. The transmit precoding matrix $\bs{W}$ and APV $\bs{T}$ are alternately optimized via a similar process as in Section \ref{Alg_MU}.
		\item \textbf{FPAs with random IRS phase shifts (FPA-RPS)}:  The $N$ MAs are deployed symmetrically to $\bs{q}_B$ and separated by the minimum distance $D_{\min}$, and the IRS reflection matrix is randomly generated similarly to the MA-RPS benchmark. The transmit precoding matrix $\bs{W}$ is optimized via the WMMSE algorithm in \cite{shi2011iteratively}.
	\end{enumerate}
	\vspace{-8pt}
	
	\subsection{Single-User System}
	\begin{figure}[t]
		\centering
		\captionsetup{justification=raggedright,singlelinecheck=false}
		\centerline{\includegraphics[width=0.45\textwidth]{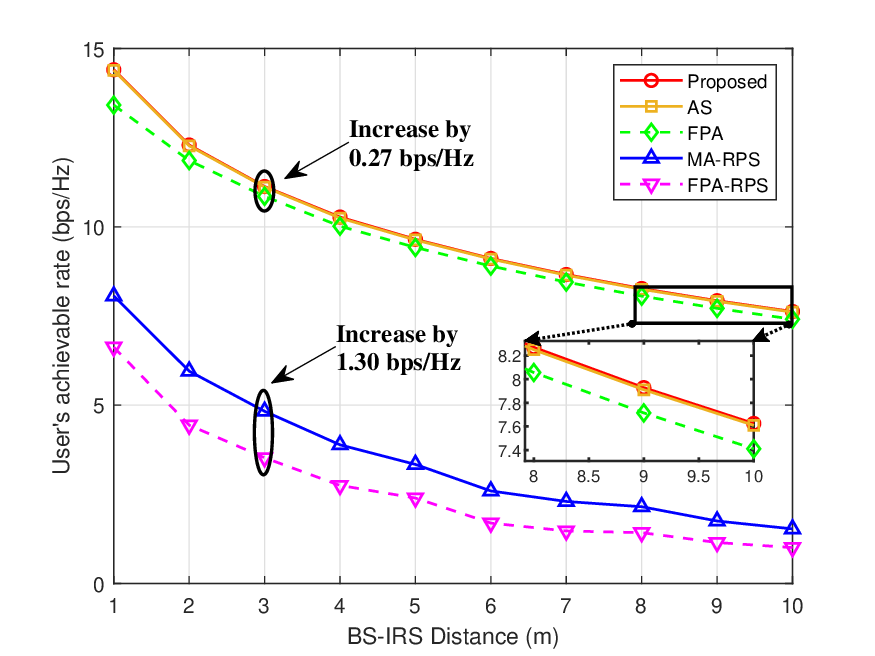}}
		\captionsetup{font=footnotesize}
		\caption{The achievable rate at the user versus the BS-IRS distance.}
		\label{Fig_SU_BI_Distance}
		\vspace{-15pt}
	\end{figure}
	
	\begin{figure}[!t]
		\centering
		\subfloat[Proposed algorithms with MAs\centering]{
			\includegraphics[width=0.45\textwidth]{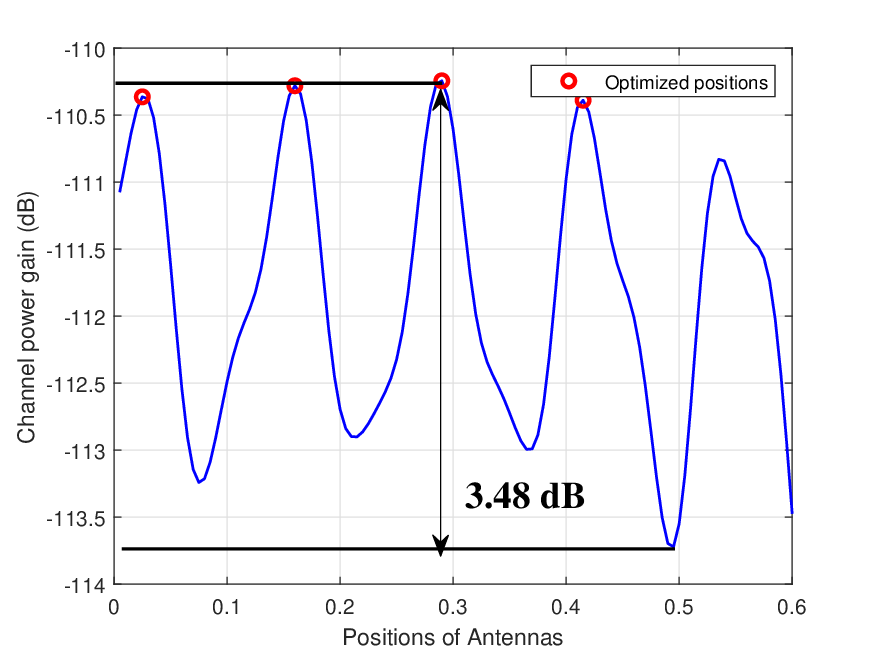}
		}
		\hfil
		\subfloat[MA-RPS benchmark\centering]{
			\includegraphics[width=0.45\textwidth]{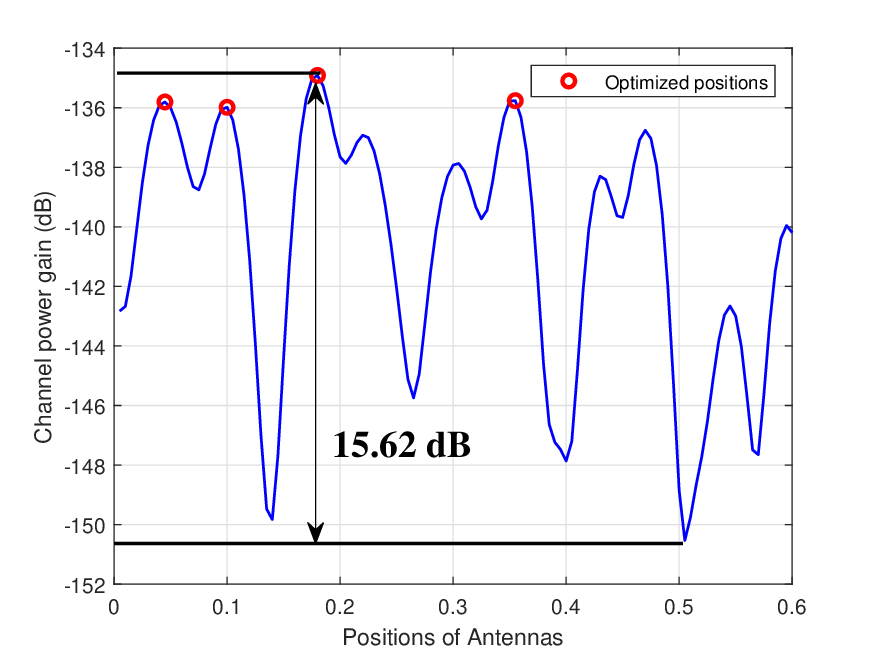}
		}
		\captionsetup{justification=raggedright,singlelinecheck=false}
		\captionsetup{font=footnotesize}
		\caption{Variation of channel power gain of the user within the transmit region.}
		\label{Fig_ChGain_SU}
		\vspace{-15pt}
	\end{figure}

	First, we consider the single-user setup and plot in Fig.~\ref{Fig_SU_BI_Distance} the achievable rate at the user versus the BS-IRS distance, with $N=4$ and $L=8$. The IRS-user distance is fixed as $d_{\text{IU}}=30$ m. It is observed that our proposed algorithm achieves higher received SNRs than all other benchmarks, thanks to channel reconfiguration abilities for both MAs and IRSs. However, the performance gain of our proposed algorithm over the FPA benchmark is marginal and decreases with the BS-IRS distance, which shows a similar trend to the case with the LoS BS-IRS channel in Fig.~\ref{Fig_LoS_SU}(b). In contrast, if the IRS phase shifts are configured randomly, employing MAs can yield more significant performance gain over the conventional FPAs. For example, when the BS-IRS distance is 3 m, the performance gain of the proposed algorithm over the FPA benchmark is only 0.27 bps/Hz (or 2.53\%); while that of the MA-RPS benchmark over the FPA-RPS benchmark is 1.30 bps/Hz (or 36.94\%). This is mainly attributed to the channel reconfiguration from the IRS. More specifically, under the optimized IRS reflection, the correlations among all transmit paths between the BS and the IRS may be decreased, which also reduces the spatial diversity gain reaped from the multi-path effect. In addition, the lack of performance gains over the FPA benchmark may also be attributed to the structural similarity between the IRS and field-response channels, as observed from \eqref{eqn_Signal} and \eqref{eqn_Channel_BI_NF}, where the path gain of each path plays a role similar to the reflection coefficient of each IRS element. This endows the field-response channel with reconfigurability, thus diminishing the performance gain by the MAs. Due to the above reasons, the performance gain of MAs over the AS benchmark is observed to be even more marginal compared to the FPA benchmark. In contrast, under the random IRS reflection, the IRS can be simply treated as a scatterer in the environment, thus barely compromising the multi-path effect.
	
	To validate the above claim, we plot in Figs.~\ref{Fig_ChGain_SU}(a) and \ref{Fig_ChGain_SU}(b) the channel power gains from the BS to the user within $\ca{C}_t$ by the proposed algorithm and the MA-RPS benchmark, respectively. It is observed that compared to the proposed algorithm, the MA-RPS benchmark yields more local maxima and minima, i.e., more significant fluctuation, within the transmit array. Furthermore, the gap between the maximum and the minimum channel power gains in the MA-RPS benchmark is around 15.62 dB, which far exceeds that in the proposed algorithm, i.e., 3.48 dB. This may be attributed to the fact that the movement region of MAs is typically limited to a wavelength level. As a result, the IRS's passive beamforming may suffice to ensure the channel power gain across all MA positions within the movement region. This effect is particularly evident in the far-field scenario, where the point-source approximation is applied to the movement region, as discussed in Section III-A. This observation validates our previous claim regarding the effects of the random versus optimized IRS reflections. \vspace{-5pt}
	
	\begin{figure}[t]
		\centering
		\captionsetup{justification=raggedright,singlelinecheck=false}
		\centerline{\includegraphics[width=0.45\textwidth]{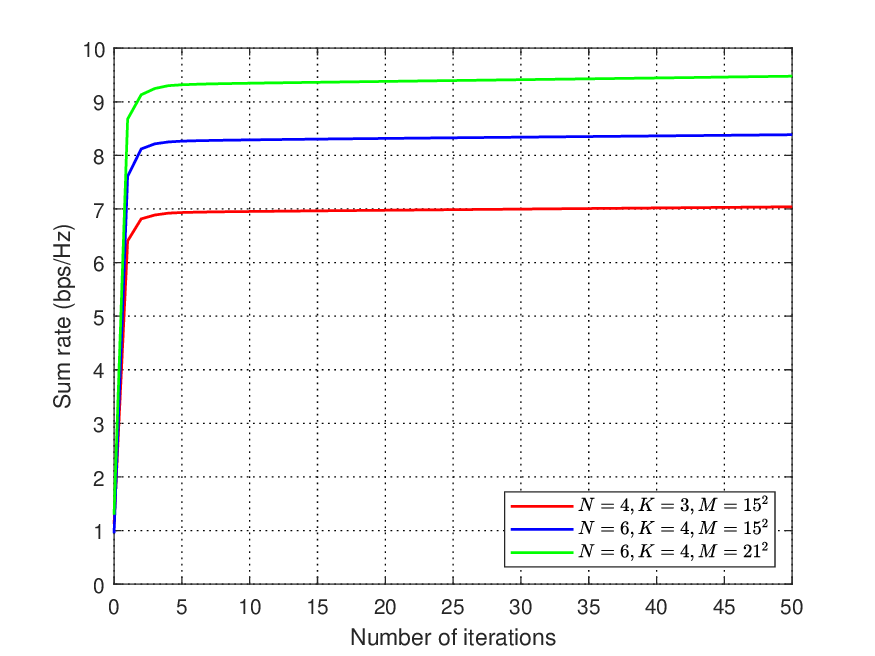}}
		\captionsetup{font=footnotesize}
		\caption{Convergence plot of the proposed algorithm.}
		\label{Fig_SumRate_Convergence}
		\vspace{-20pt}
	\end{figure}
	
	\subsection{Multi-User System}
	In this subsection, we present the simulation results for the multi-user system setup. First, we plot in Fig.~\ref{Fig_SumRate_Convergence} the convergence of our proposed algorithm in Section IV under three different setups. It is observed that the sum-rate of our proposed algorithm monotonically increases with the iteration number and converges after about only 5 iterations for all setups considered, which manifests the efficiency of our proposed algorithm.

	\begin{figure}[t]
		\centering
		\captionsetup{justification=raggedright,singlelinecheck=false}
		\centerline{\includegraphics[width=0.45\textwidth]{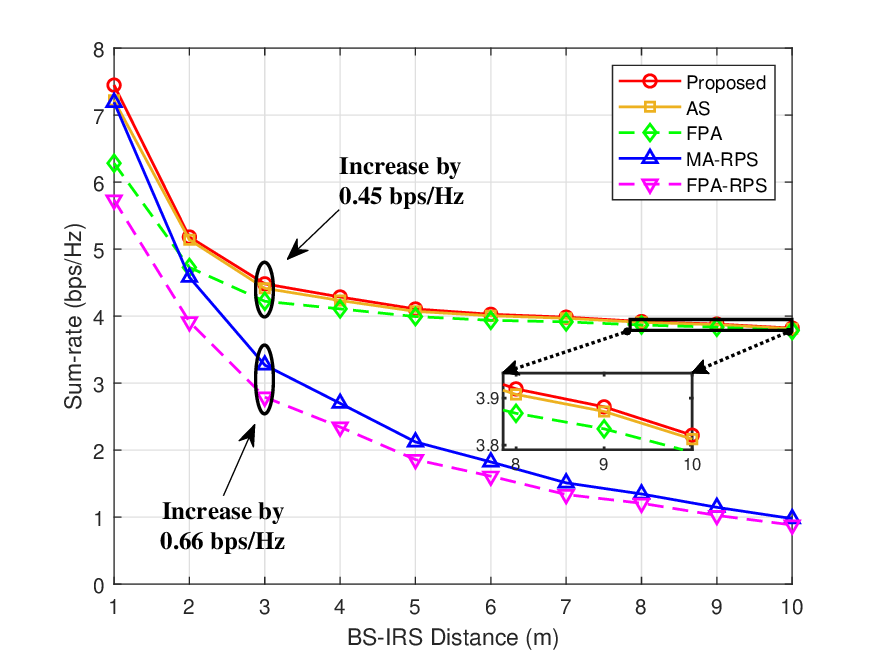}}
		\captionsetup{font=footnotesize}
		\caption{Sum-rate versus the BS-IRS distance under the LoS BS-IRS channel.}
		\label{Fig_SumRate_BS-IRS_Distance_LoS}
		\vspace{-15pt}
	\end{figure}
	
	\begin{figure}[!t]
		\centering
		\subfloat[Proposed algorithms with MAs\centering]{
			\includegraphics[width=0.45\textwidth]{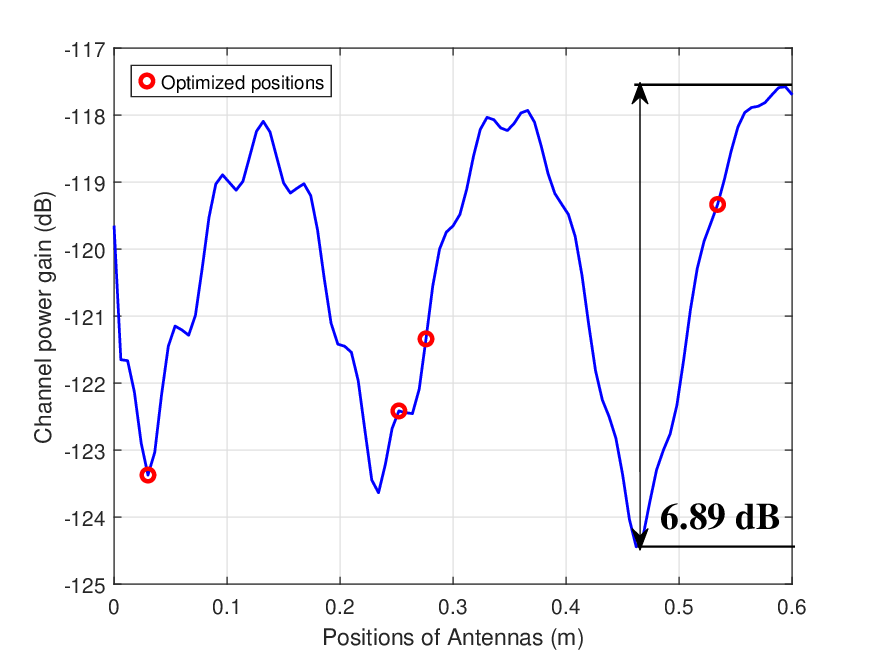}
		}
		\hfil
		\subfloat[MA-RPS benchmark\centering]{
			\includegraphics[width=0.45\textwidth]{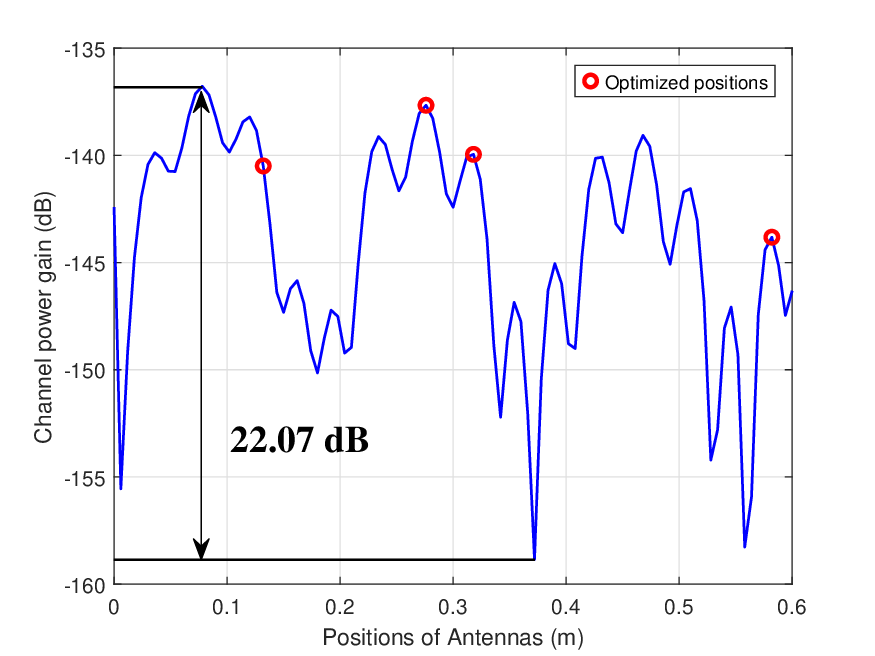}
		}
		\captionsetup{justification=raggedright,singlelinecheck=false}
		\captionsetup{font=footnotesize}
		\caption{Variation of channel power gain of user 1 within the transmit region.}
		\label{Fig_ChGain}
		\vspace{-15pt}
	\end{figure}
	
	In Fig.~\ref{Fig_SumRate_BS-IRS_Distance_LoS}, we plot the sum rates by different schemes versus the BS-IRS distance under the LoS  BS-IRS channel model. It is observed from Fig.~\ref{Fig_SumRate_BS-IRS_Distance_LoS} that the sum rates by all considered schemes decrease as the BS-IRS distance increases due to the more severe path loss. Nonetheless, the IRS plays a more significant role in affecting the sum-rate performance compared to MA, as inferred from the larger performance gap between FPA and FPA-RPS than that between MA-RPS and FPA-RPS. Besides, the performance gain of the proposed algorithm over the FPA benchmark is observed to decrease rapidly with the BS-IRS distance. This is consistent with our analysis in Section IV-A, as the BS-IRS channel is closer to the far-field region as the BS-IRS distance increases. Hence, we can conclude that for the LoS BS-IRS channel, employing MAs cannot yield significant performance gain over FPAs in general for both single-user and multi-user systems in the presence of the optimized IRS reflection. Moreover, compared with the optimized IRS reflection, it can be seen that the MAs can yield more performance gain under the random IRS reflection, e.g., 0.45 bps/Hz (or 6.17\%) versus 0.66 bps/Hz (or 17.16\%) at 3 m, which is similar to the observation made from Fig.~\ref{Fig_SU_BI_Distance}. In Figs.~\ref{Fig_ChGain}(a) and~\ref{Fig_ChGain}(b), we also plot the channel power gains from the BS to user 1 within $\ca{C}_t$ by the proposed algorithm and the MA-RPS benchmark, respectively. It is observed that the gap between the maximum and the minimum channel power gains in the MA-RPS benchmark is around 22.07 dB, which is much larger than that in the proposed algorithm, i.e., 6.89 dB. It is also interesting to note that more antenna positions are optimized as local maxima in Fig.~\ref{Fig_ChGain}(b) compared to Fig.~\ref{Fig_ChGain}(a). Particularly, the position of the first antenna is even optimized as a local minimum point in Fig.~\ref{Fig_ChGain}(a) for multi-user interference mitigation. This implies that the MA position optimization places greater emphasis on mitigating multi-user interference under optimized IRS reflection compared to random IRS reflection.
	
	\begin{figure}[t]
		\centering
		\captionsetup{justification=raggedright,singlelinecheck=false}
		\centerline{\includegraphics[width=0.45\textwidth]{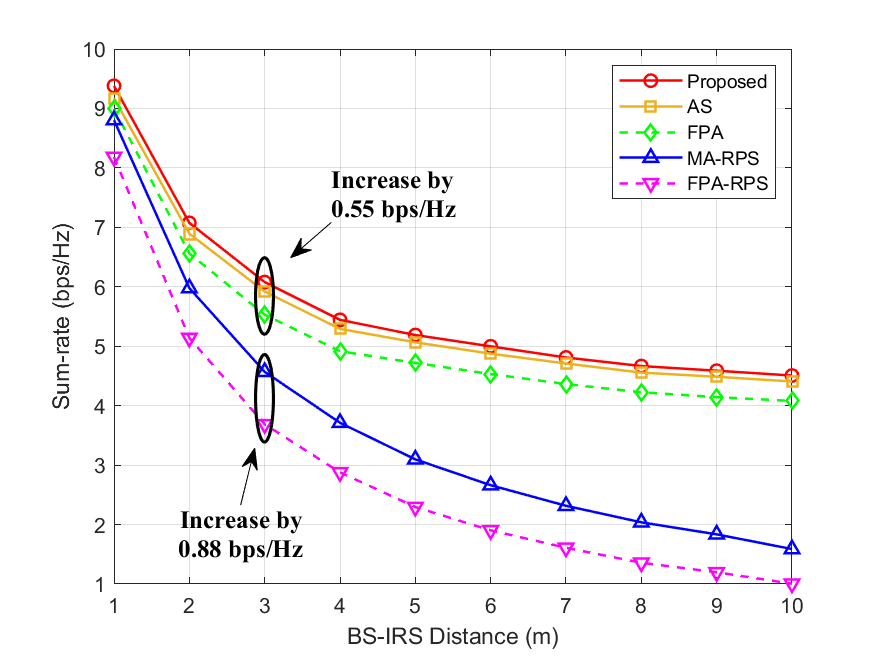}}
		\captionsetup{font=footnotesize}
		\caption{Sum-rate versus the BS-IRS distance under the multi-path BS-IRS channel.}
		\label{Fig_SumRate_BS-IRS_Distance_MultiPath}
		\vspace{-15pt}
	\end{figure}
	
	Next, we plot in Fig.~\ref{Fig_SumRate_BS-IRS_Distance_MultiPath} the sum rates by different schemes versus the BS-IRS distance under the multi-path BS-IRS channel model with $L=4$. Unlike the observations made from Fig.~\ref{Fig_SumRate_BS-IRS_Distance_LoS}, it is observed from Fig.~\ref{Fig_SumRate_BS-IRS_Distance_MultiPath} that even with a large BS-IRS distance, the performance gap between MAs and FPAs still exists. In particular, the performance gap first increases with the BS-IRS distance and remains approximately static for both optimized and random IRS phase shifts. The above observation implies that MAs have a more significant effect on the sum-rate performance under the multi-path BS-user channel model. Moreover, the performance gap between MAs and FPAs becomes more significant under random IRS phase shifts than optimized IRS phase shifts, e.g., 0.88 bps/Hz versus 0.55 bps/Hz as the BS-IRS distance is 3 m. 
	
	\begin{figure}[t]
		\centering
		\captionsetup{justification=raggedright,singlelinecheck=false}
		\centerline{\includegraphics[width=0.45\textwidth]{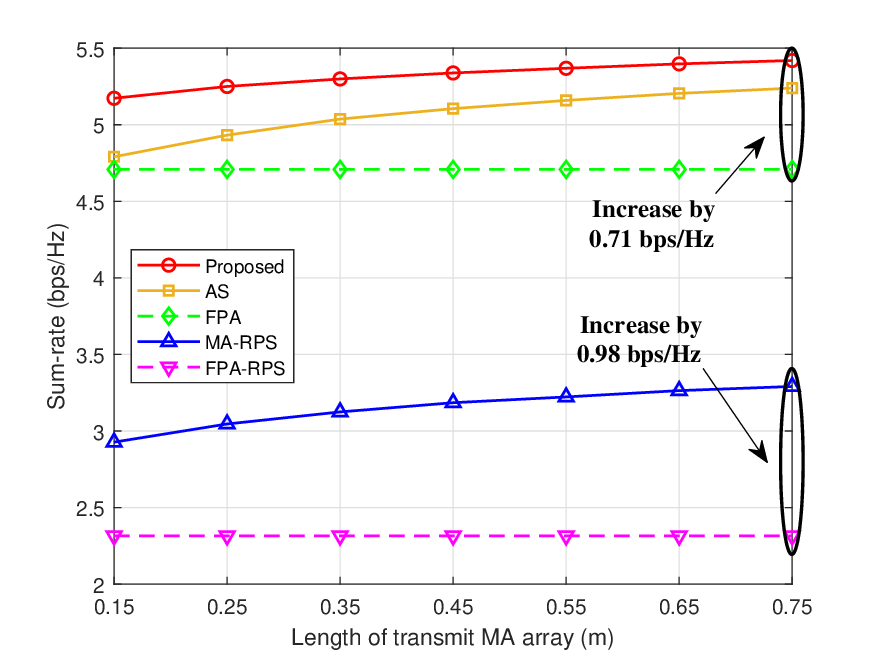}}
		\captionsetup{font=footnotesize}
		\caption{Sum-rate versus the length of the transmit array.}
		\label{Fig_SumRate_TxRegion}
		\vspace{-15pt}
	\end{figure}
	\begin{figure}[t]
		\centering
		\captionsetup{justification=raggedright,singlelinecheck=false}
		\centerline{\includegraphics[width=0.45\textwidth]{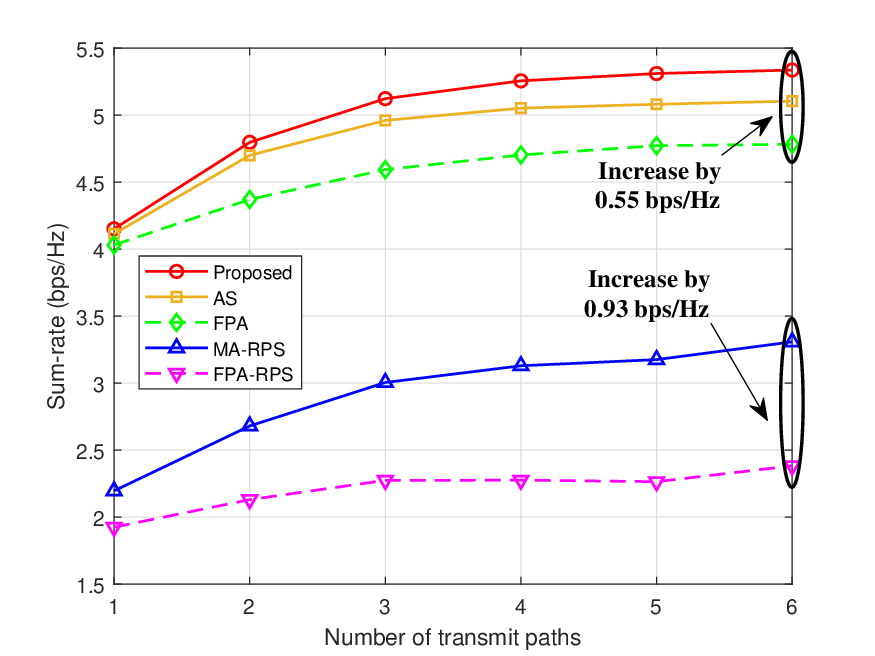}}
		\captionsetup{font=footnotesize}
		\caption{Sum-rate versus the number of transmit paths.}
		\label{Fig_SumRate_PathNums}
		\vspace{-15pt}
	\end{figure}
	
	In Fig.~\ref{Fig_SumRate_TxRegion}, we plot the sum rates by different schemes versus the length of the transmit array $\ca{C}_t$, i.e., $A$. It is observed that the sum-rates by both the proposed algorithm and the MA-RPS benchmark increase with the length of $\ca{C}_t$. This is because a larger transmit region offers more spatial degrees of freedom (DoFs) for the MAs to improve the multi-user performance. Moreover, the performance gain of the proposed algorithm over the FPA benchmark is observed to increase slowly with $A$, which is possibly due to the limited spatial diversity in the presence of the optimized IRS passive beamforming. This claim is also supported by the observation that the performance gain under random IRS reflection is higher than that under optimized IRS reflection, e.g., 0.98 bps/Hz (or 42.15\%) versus 0.71 bps/Hz (or 15.11\%) at 0.75 m.
	
	Lastly, in Fig.~\ref{Fig_SumRate_PathNums}, we plot the sum-rate performance versus the number of transmit paths between the BS and the IRS. Note that in the case of a single transmit path, it is equivalent to the LoS BS-IRS channel. It is observed that the sum-rate performance by all schemes (except the FPA-RPS benchmark) increases with the number of transmit paths, thanks to the more significant small-scale fading. It can be inferred that MAs are more preferred if there are a large number of scatterers between the BS and the IRS, such that relying on the IRS alone cannot fully reconfigure the BS-IRS channel as desired. However, when there only exists a single transmit path, the performance gain of the MAs over the FPAs is negligible with both optimized and random IRS reflection, which is consistent with our analysis in Section IV-A. 
	
	\section{Conclusions and Future Directions}
	In this paper, we investigated a joint active/passive beamforming and BS antenna position optimization problem for an IRS-assisted multi-user MISO MA system. To gain useful insights, we first conducted performance analysis and solved the SNR maximization problem in the single-user case. Then, we solved the sum-rate maximization problem using the AO algorithm in the multi-user case. Both of our analytical and numerical results revealed that the presence of an IRS can hinder the performance gain of the MAs over conventional FPAs if the BS-user direct link is blocked. The main takeaways are summarized as follows. First, for IRS-assisted MA systems, if the BS-IRS channel is LoS-dominated and the IRS is configured with an optimized reflection, the performance gain of MAs over FPAs diminishes with the BS-IRS distance and ultimately vanishes. Second, if the BS-IRS channel has non-negligible multi-path components, MAs generally yield a more significant gain over FPAs compared to the LoS-dominated BS-IRS channel, under both near- and far-field propagation conditions. Third, the MAs tend to yield a more significant gain over FPAs if the IRS is configured with a random reflection instead of an optimized one.
	
	This paper can be extended to various directions as future work. First, we only consider the cascaded BS-IRS-user channel, ignoring the direct BS-user channel. In the presence of the direct BS-user channel, the performance gain of MAs depends on more factors such as the strength ratio of the direct link to the reflected link, as well as the characteristics of the individual BS-user and BS-IRS channels, which renders the performance analysis more challenging and deserves further in-depth investigation. Second, we only consider BS-side MAs in this paper, and their positions need to cater to all users via only their shared BS-IRS channel. Hence, it is worthy of investigating the performance gain of user-side MAs over FPAs, which may have a larger DoF for performance enhancement by reconfiguring their IRS-user channels individually. Third, we assume perfect CSI on all links to focus on the performance analysis and optimization in this paper. However, the coexistence of the IRS and MAs may significantly increase the CSI estimation overhead. It is thus interesting to study whether these two technologies can be reciprocal in terms of channel estimation. Last but not least, other more general system setups, e.g., MIMO and/or multi-reflection-IRS systems, can also be studied to explore the interactions between MAs and multiple IRSs.
	
	\bibliography{MA_IRS_Ref.bib}
	\bibliographystyle{IEEEtran}
\end{document}